\documentclass[a4paper,11pt]{article}
\pdfoutput=1
\usepackage{jheppub}
\usepackage[T1]{fontenc}
\usepackage{graphicx}
\usepackage{multirow}
\usepackage{xcolor}
\usepackage{filecontents}
\usepackage[normalem]{ulem} 

\begin{document}
\title{\boldmath Phenomenological Modeling of the $^{163}$Ho Calorimetric Electron Capture Spectrum from the HOLMES Experiment}

\author[a,b]{F. Ahrens,}
\author[c]{B. K. Alpert,}
\author[d]{D. T. Becker,}
\author[c]{D. A. Bennett,}
\author[a,b,e]{E. Bogoni,}
\author[e,f]{M. Borghesi,}
\author[e,f]{P. Campana,}
\author[e,f]{R. Carobene,}
\author[e,f]{A. Cattaneo,}
\author[a,b]{A. Cian,}
\author[e,f]{H. A. Corti,}  
\author[a,b]{N. Crescini,}
\author[g]{M. De Gerone,}
\author[c]{W. B. Doriese,}
\author[e,f]{M. Faverzani,}
\author[g,h]{L. Ferrari Barusso,}
\author[f]{E. Ferri,}
\author[c]{J. Fowler,}
\author[g]{G. Gallucci,}
\author[e,f]{S. Gamba,}
\author[d]{J. D. Gard,}
\author[i,j,k]{H. Garrone,}
\author[g,h]{F. Gatti,}
\author[e,f]{A. Giachero,}
\author[e,f]{M. Gobbo,}
\author[a,b,e]{A. Irace,}
\author[l]{U. K\"oster,}
\author[e,f]{D. Labranca,}
\author[m,n]{M. Lusignoli,}
\author[i,j,o]{F. Malnati,}
\author[a,b]{F. Mantegazzini,}
\author[a,b]{B. Margesin,}
\author[c]{J. A. B. Mates,}
\author[p]{E. Maugeri,}
\author[b,q]{R. Mezzena}
\author[i,j]{E. Monticone,}
\author[e,f]{R. Moretti,}
\author[e,f]{A. Nucciotti \note{https://orcid.org/0000-0002-8458-1556},}
\author[c]{G. C. O'Neil,}
\author[f]{L. Origo,}
\author[f]{G. Pessina,}
\author[e,f]{S. Ragazzi,}
\author[i,j]{M. Rajteri,}
\author[c]{C. D. Reintsema,}
\author[c]{D. R. Schmidt,}
\author[c]{D. S. Swetz,}
\author[p]{Z. Talip,}
\author[c,d]{J. N. Ullom,}
\author[c]{and L. R. Vale}

\affiliation[a]{Fondazione Bruno Kessler (FBK), Via Sommarive 18, Trento, Italy}
\affiliation[b]{Trento Institute for Fundamental Physics and Applications (TIFPA), INFN, Via Sommarive 18, Trento, Italy}
\affiliation[c]{Quantum Sensor Division, National Institute of Standards and Technology (NIST), 325 Broadway, Boulder, CO, USA}
\affiliation[d]{Department of Physics, University of Colorado, 2000 Colorado Ave, Boulder, CO, USA}
\affiliation[e]{Dipartimento di Fisica, Universit\`a di Milano-Bicocca, Piazza della Scienza 3, Milano, Italy}
\affiliation[f]{Istituto Nazionale di Fisica Nucleare (INFN), Sezione di Milano-Bicocca, Piazza della Scienza 3, Milano, Italy}
\affiliation[g]{Istituto Nazionale di Fisica Nucleare (INFN), Sezione di Genova, Via Dodecaneso 33, Genova, Italy}
\affiliation[h]{Dipartimento di Fisica, Universit\`a di Genova, Via Dodecaneso 33, Genova, Italy}
\affiliation[i]{Istituto Nazionale di Ricerca Metrologica (INRiM), Strada delle Cacce 91, Torino, Italy}
\affiliation[j]{Istituto Nazionale di Fisica Nucleare (INFN), Sezione di Torino, Via Pietro Giuria 1, Torino, Italy}
\affiliation[k]{Dipartimento di Elettronica e Telecomunicazioni, Politecnico di Torino, Corso Duca degli Abruzzi 24, Torino, Italy}
\affiliation[l]{Institut Laue-Langevin (ILL), 71 avenue des Martyrs, Grenoble, France}
\affiliation[m]{Dipartimento di Fisica, Sapienza Universit\`a di Roma, Piazzale Aldo Moro 2, Roma, Italy}
\affiliation[n]{Istituto Nazionale di Fisica Nucleare (INFN), Sezione di Roma 1, Piazzale Aldo Moro 2, Roma, Italy}
\affiliation[o]{Dipartimento Scienza Applicata e Tecnologia, Politecnico di Torino, Corso Duca degli Abruzzi 24, Torino, Italy}
\affiliation[p]{PSI Center for Nuclear Engineering and Sciences, Forschungsstrasse 111, 5232 Villigen PSI, Switzerland}
\affiliation[r]{Dipartimento di Fisica, Universit\`a di Trento, Via Sommarive 14, Trento, Italy}

\emailAdd{angelo.nucciotti@mib.infn.it}
\emailAdd{dan.schmidt@nist.gov}
\abstract{
We present a comprehensive phenomenological analysis of the calorimetric electron capture (EC) decay spectrum of $^{163}$Ho as measured by the HOLMES experiment. Using high-statistics data, we unfold the instrumental energy resolution from the measured spectrum and model it as a sum of Breit-Wigner resonances and shake-off continua, providing a complete set of parameters for each component.
Our approach enables the identification and tentative interpretation of all observed spectral features, including weak and overlapping structures, in terms of atomic de-excitation processes. We compare our phenomenological model with recent {\it ab initio} theoretical calculations, finding good agreement for both the main peaks and the spectral tails, despite the limitations of current theoretical and experimental precision. The model delivers an accurate description of the endpoint region, which is crucial for neutrino mass determination, and allows for a realistic treatment of backgrounds such as pile-up and tails of low-energy components. Furthermore, our decomposition facilitates the generation of Monte Carlo toy spectra for sensitivity studies and provides a framework for investigating systematic uncertainties related to solid-state and detector effects. This work establishes a robust foundation for future calorimetric neutrino mass experiments employing $^{163}$Ho, supporting both data analysis and experimental design.
}
\maketitle
\section{Introduction}
\label{sec:holmes}
The measurement of the absolute neutrino mass has far-reaching implications in terms of its impact on cosmology, particle physics, and our understanding of the dynamics of the universe \cite{divalentino_most_2021}.
Direct neutrino mass experiments, such as KATRIN\,\cite{katrincollaboration_direct_2025}, determine the absolute neutrino mass scale by precisely measuring the kinematics of low-$Q$ beta decays near their end-point. This approach is both sensitive and largely model-independent, relying solely on energy and momentum conservation. HOLMES is a project that aims to directly measure the electron neutrino mass $m_{\nu}$ using calorimetric methods to study the end point of electron-capture (EC) decay of $^{163}$Ho \cite{alpert_holmes_2015b}. 
A similar approach is pursued by the ECHo (Electron Capture in $^{163}$Ho) experiment\,\cite{gastaldo_electron_2017}, which also investigates the EC spectrum of $^{163}$Ho to probe the neutrino mass.

In an ideal calorimetric experiment\,\cite{nucciotti_use_2016b}, the radioactive source is fully embedded within the detector, ensuring that only the neutrino escapes undetected. All other energy released in the decay --including that associated with atomic or molecular excitations-- is measured via the prompt de-excitation of these states, provided their lifetimes are short compared to the detector's response time. Consequently, the observable in such an experiment is the total energy deposited in the detector, which corresponds to the decay $Q$-value minus the neutrino energy, $E_\nu$.

\begin{figure*}[tbh]
  \centering
  \includegraphics[width=0.8\textwidth]{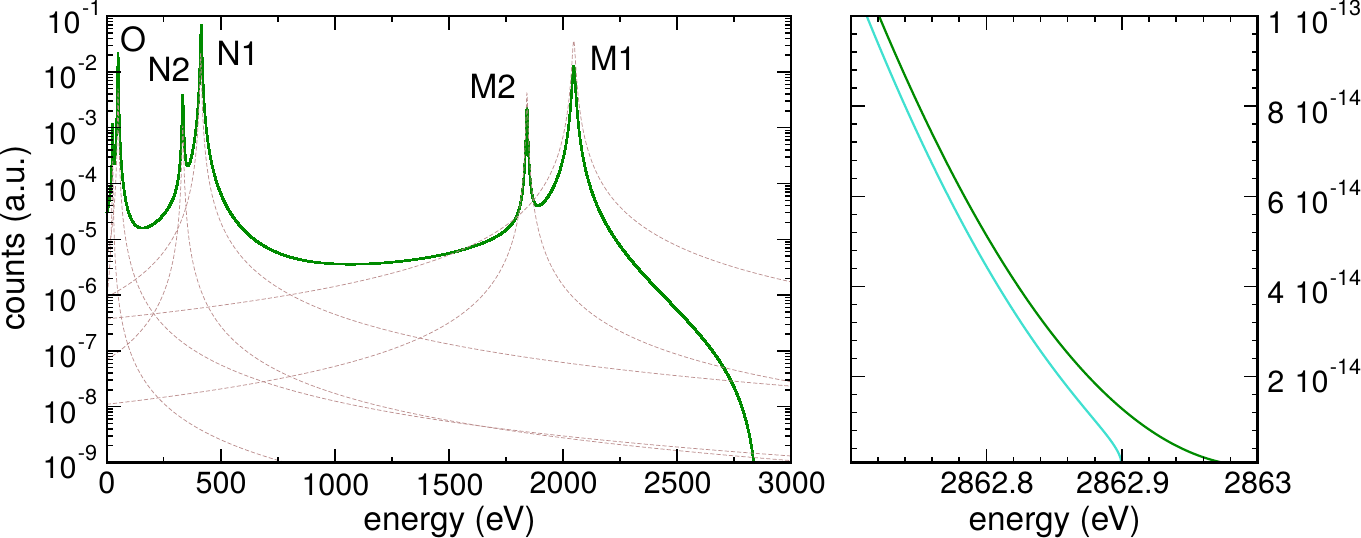}
\caption{\label{fig:spe1H}Calorimetric EC decay spectrum in the single-hole approximation, as described in \cite{derujula_calorimetric_1982}. Dashed lines indicate the individual Breit-Wigner resonances from Eq.\,\ref{eq:E_c-distr}, shown without the phase space factor, ({\it i.e.}, omitting the prefactor before the summation in Eq.\,(\ref{eq:E_c-distr})). The right panel displays a zoom of the end-point region for $m_{\nu} = 0$\,eV (green) and $m_{\nu} = 0.1$\,eV (cyan).}
\end{figure*}

The concept of using calorimetric measurements to determine the neutrino mass was first introduced by J.J.\,Simpson \cite{simpson_measurement_1981}. He recognized that embedding the radioactive source within the detector could mitigate systematic uncertainties arising from atomic and molecular final states, as well as from electron interactions in external-source spectrometric experiments, such as those employing tritium. Building on this idea, A.\,De Rujula and M.\,Lusignoli subsequently proposed the use of calorimetry with $^{163}$Ho  \cite{derujula_calorimetric_1982}, an electron-capture isotope with an extremely low $Q$-value, thereby maximizing sensitivity to the neutrino mass.
When A. De R\'ujula and M. Lusignoli first proposed the calorimetric measurement of the EC decay of $^{163}$Ho \cite{derujula_calorimetric_1982}, they employed Fermi's golden rule to compute the distribution of the de-excitation (calorimetric) energy $E_c$. This approach models the spectrum as a sum of Breit-Wigner lines, each with a natural width $\Gamma_i$ centered at the ionization energies $E_i$ ({\it i.e.}, the binding energies) of the captured electrons. For a non-zero $m_{\nu}$, the resulting distribution is
\begin {eqnarray}
\label{eq:E_c-distr}
N(E_c,m_{\nu}) = {G_{\beta}^2 \over {8 \pi^3}}(Q-E_c) \sqrt{(Q-E_c)^2-m_{\nu}^2} \sum_i n_i  C_i \beta_i^2 B_i {\Gamma_i \over (E_c-E_i)^2+\Gamma_i^2/4}
\end{eqnarray}
where $G_{\beta} = G_F \cos \theta_C$ (with the Fermi constant $G_F$ and the Cabibbo angle $\theta_C$),
$n_i$  is the fraction of occupancy, $C_i $ is the nuclear shape factor,
$\beta_i$ is the Coulomb amplitude of the electron radial wave function ({\it i.e.} the modulus of the wave function at the origin)
and $B_i$ is an atomic correction for electron exchange and overlap \cite{faessler_electron_2015}.
The sum in (\ref{eq:E_c-distr}) runs over final states in which the daughter Dy atom has a vacancy in one of the atomic shells that are energetically accessible for electron capture, given the transition energy $Q = (2863.2\pm0.6)$\,eV \cite{schweiger_penning-trap_2024}. Only shells for which the Coulomb amplitude $\beta_i$ is not negligible contribute significantly, namely M1 ($3s$), M2 ($3p_{1/2}$), N1 ($4s$), N2 ($4p_{1/2}$), O1 ($5s$), O2 ($5p_{1/2}$), and P1 ($6s$). The resulting spectrum is shown in Fig.\,\ref{fig:spe1H}, where the individual contributions from each shell are also indicated.

\subsection{Aim of the present work}
In this work, we present the first high-statistics calorimetric spectrum of the EC decay of $^{163}$Ho as measured by the HOLMES experiment, and compare it with current theoretical predictions to advance the understanding of this process. After introducing the EC decay and summarizing the relevant theoretical developments for $^{163}$Ho, we provide a phenomenological and analytical description of the measured spectrum. Our objectives are twofold: first, to relate the phenomenological parameters to spectral features predicted by different theoretical models; and second, to establish a robust analytical framework suitable for both neutrino mass extraction from the HOLMES data and for evaluating the statistical sensitivity of future holmium-based neutrino mass experiments.

\subsection{Beyond the single hole approximation}
\label{ssec:beyond}
It was recognized early on that if EC is properly considered as a process involving the whole atom, more atomic excitations are expected in addition to the single hole ones included in (\ref{eq:E_c-distr}) ({\it e.g.}, \cite{bambynek_orbital_1977}). With the sudden change of the nuclear charge caused by the EC, the wave functions of the Dy atomic orbitals overlap only partially with those of parent Ho. Moreover, the neutral daughter Dy atom has an extra (the eleventh) $4f$ electron, because the parent Ho atom has an electronic configuration which differs from the one of Dy in the number of $4f$ electrons (11 vs. 10) (see also \cite{springer_enhanced_1985}).

\begin{figure*}[tbh]
  \centering
  \includegraphics[width=0.325\textwidth]{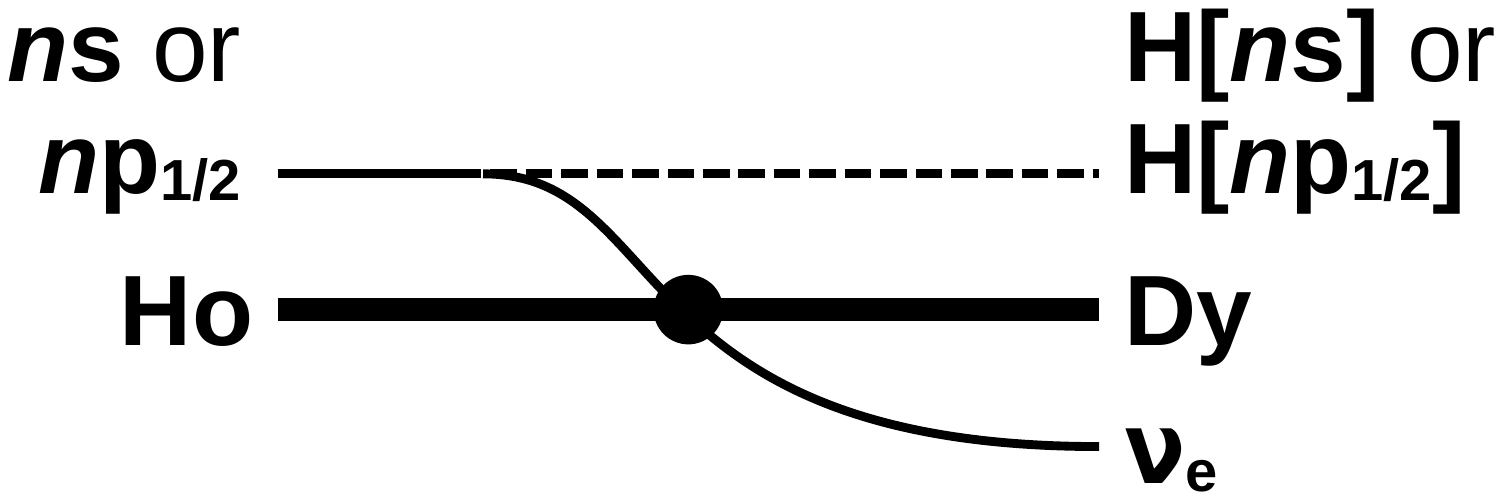}
   \includegraphics[width=0.325\textwidth]{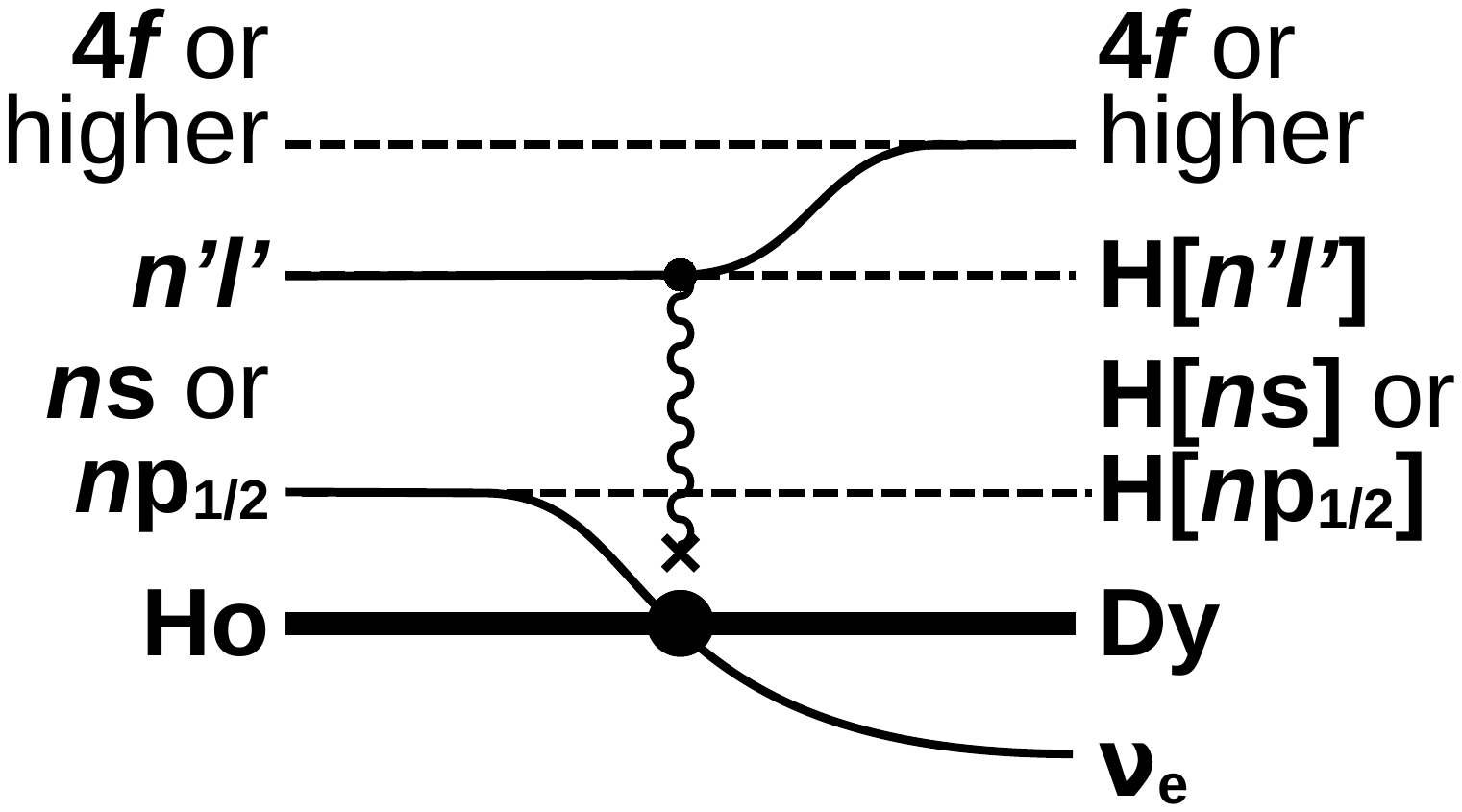}
    \includegraphics[width=0.325\textwidth]{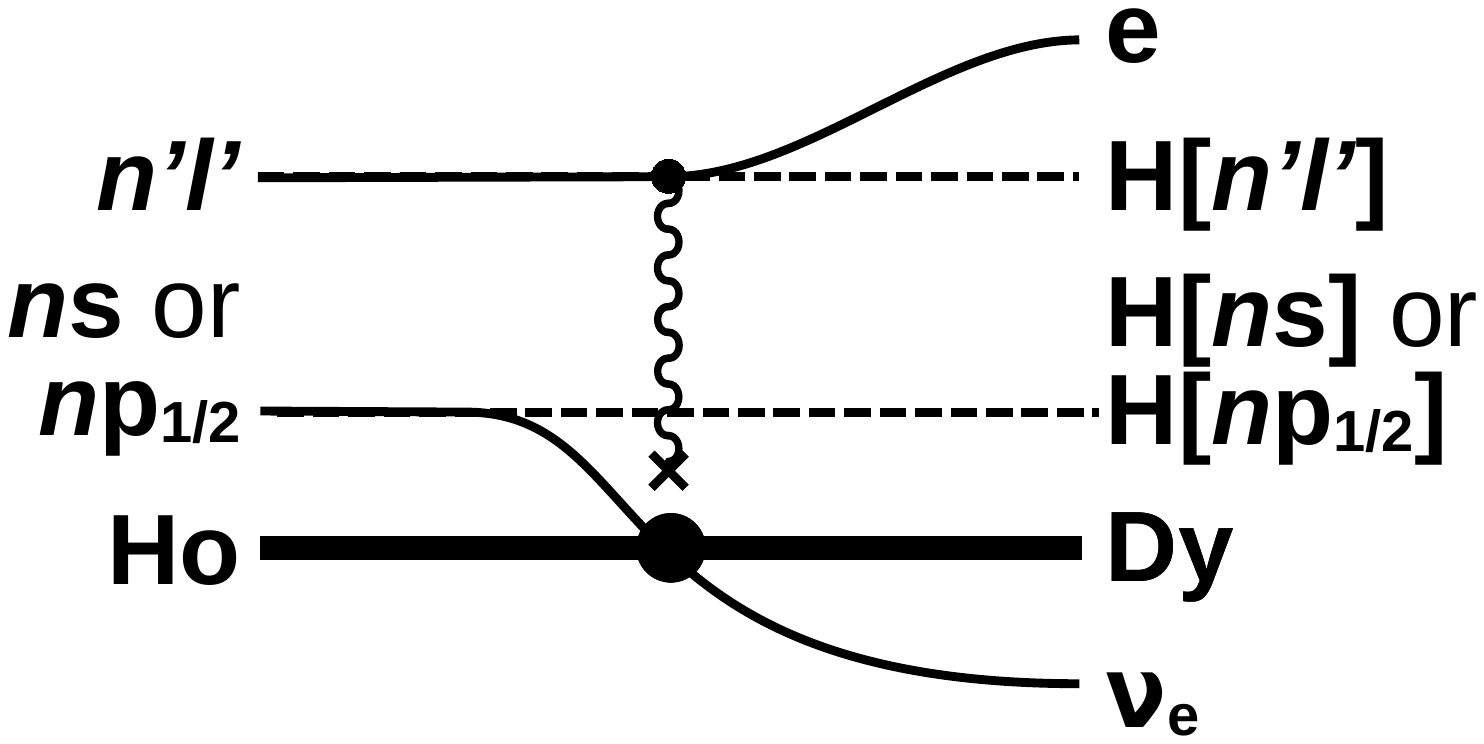}
\caption{Schematic representation of atomic excitations following EC: single-hole creation (left), double-hole via shake-up (center), and double-hole via shake-off (right). Here, $n=4,5,6$ denotes the principal quantum numbers of the core holes.}
  \label{fig:diagr}
\end{figure*}
Including the atomic degrees of freedom in the initial and final states used to calculate the EC transition probability, additional excitations of the final Dy atom are expected.
Atomic transitions occurring {\it during} the EC process involve the neutrino in reactions that can be 2-body, 3-body, or of even higher order. The final state consists of the excited Dy daughter atom, the neutrino, and, in some cases, unbound electrons or photons.
Energy conservation requires $Q-m_\nu=E_c+E_\nu$, where $E_c$ now includes all the atomic excitations of the Dy daughter atom and the energy of electrons or photons eventually emitted (and absorbed in the detector).

In 2-body processes, the captured electron scatters via Coulomb interaction with one of the spectator electrons in the same or higher shell of the Ho atom, raising it to an unoccupied bound shell (Fig.\,\ref{fig:diagr}, center). These so called shake-up processes, leave
the neutral Dy daughter atom with two holes and two displaced electrons, one being in the 4$f$ shell:
one hole, H1, is caused by the capture of an electron in shell M1 or above, and the other, H2, is the one left  by the {\it  shaken} electron in a shell above that of H1.
After the EC, the atomic de-excitation releases in the detector the binding energies of the two missing electrons, contributing to the $E_c$ spectral distribution with a peak at $B[\mathrm{H1}]+B[\mathrm{H2}]$, where the $B[\mathrm{H}]$s must be evaluated for the final state orbital configuration that is the one of a neutral Dy atom with an extra $4f$ electron. The peak is expected again to have a Breit-Wigner shape with a $\Gamma$ which is the sum of the line-widths $\Gamma_1$ and $\Gamma_2$ given by the lifetime of each of the two vacancies H1 and H2, respectively.

Two types of 3-body processes can occur during EC (see again, for example, \cite{bambynek_orbital_1977}): the internal bremsstrahlung or radiative EC, in which the captured electron emits a photon, and the internal ionization or shake-off, in which the captured electron scatters the spectator electron into an unbound state  (Fig.\,\ref{fig:diagr}, right). In both cases, the emitted particle has a continuous spectrum of energies and energy conservation is enforced including the neutrino in the balance. In the case of $^{163}$Ho, shake-off is estimated to be the most relevant of the two processes \cite{bambynek_orbital_1977,brass_initio_2021a,derujula_calorimetric_2016}.
As in the shake-up case, shake-off processes leave the ionized Dy atom with two holes and an extra $4f$ electron, as well as an unbound electron. Consequently, they introduce additional broad structures in the $E_c$ distribution (\ref{eq:E_c-distr}), extending from the sum of the binding energies of the two missing electrons, $B[\mathrm{H1}]+B[\mathrm{H2}]$, up to $Q-m_\nu$.

\subsection{Previous theoretical work and experimental data}
\label{ssec:previous}
Shake-up and shake-off processes are not unique to EC but have also been observed in other processes where the charge of the nucleus changes, such as in beta decay \cite{migdal_ionization_1941,carlson_calculation_1968},  or where a hole in a core atomic shell is suddenly created, including internal conversion \cite{robertson_shakeup_2020}, X-ray photoelectron spectroscopy (XPS) \cite{carlson_calculation_1973}, and X-ray atomic physics \cite{dean_initio_2024}.

In EC, shake processes are generally expected to be less probable than in other processes due to the partial compensation between the opposite charge changes occurring in the nucleus and in the captured atomic shell \cite{crasemann_atomic_1979}.
Nevertheless, the renewed interest in neutrino mass experiments exploiting the calorimetric measurement of the $^{163}$Ho EC decay spectrum and the publication of first spectra measured by the ECHo experiment starting from 2012 \cite{ranitzsch_development_2012} have prompted many authors to investigate the effect of higher order excitations on the spectral shape and, in particular, on the end-point region \cite{alpert_most_2025,brass_initio_2020,brass_initio_2018,faessler_neutrino_2017,derujula_calorimetric_2016,robertson_examination_2015}.
Initial ECHo data \cite{gastaldo_characterization_2013,ranitzsch_development_2012a,hassel_recent_2016,ranitzsch_first_2014} clearly revealed that model (\ref{eq:E_c-distr}) was inadequate. Consequently, various authors pursued enhancements to (\ref{eq:E_c-distr}) through different approaches. De R\'ujula in \cite{derujula_two_2013} proposed improvements to his seminal work \cite{derujula_calorimetric_1982} incorporating quantum interferences to alter resonance shapes and ``instantaneous'' electron ejection (shake processes). He estimated the double hole excitation rate, finding it insignificant for the spectrum's end-point shape. Contrarily, Robertson \cite{robertson_examination_2015} and Faessler {\it et al.} \cite{faessler_improved_2015} challenged this, providing more accurate shake process probabilities derived from Carlson and Nestor theory for Xe \cite{carlson_calculation_1973} and from a relativistic Dirac-Hartree-Fock (DHF) approach with Slater determinants calculated for Ho and Dy, respectively.
Despite only limited agreement in their probability estimates, both approaches predict a number of excitations with one core hole caused by the EC (M1, M2, N1, N2, ...) and a second in an outer shell, implying that the calorimetric spectra should exhibit multiple secondary peaks on the high-energy wings of the core resonances in (\ref{eq:E_c-distr}).
However, the theoretical spectra significantly deviate from the preliminary experimental results \cite{ranitzsch_development_2012, croce_development_2016, hassel_recent_2016}, which, despite low statistics, display only broad features alongside the single hole resonances.
To refine understanding, new studies from  De R\'ujula {\it et al.} \cite{derujula_calorimetric_2016} and Faessler {\it et al.} \cite{faessler_neutrino_2017} have incorporated the broad spectral contributions from shake-off processes, again adopting different approaches.
While a detailed comparison of these methodologies is beyond the scope of this paper, it is noteworthy that the approach in \cite{derujula_calorimetric_2016} appears to provide a better fit to experimental data, although it requires some empirical adjustment of the shake-off intensities, particularly for the broad satellites of the N1 resonance.
However, a common conclusion of all these works is the prediction of several two-hole excitations ({\it i.e.}, shake-up and shake-off) which are not seen in the experimental data.

A novel approach has been recently adopted by Brass {\it et al.} \cite{brass_initio_2018,brass_initio_2020}, prompted by challenges in explaining ECHo's high-resolution spectrum \cite{velte_high-resolution_2019}.
They employ what is termed an \textit{ab initio} strategy, which involves the first-principles computation of resonance energies $E_i$ and their associated broadenings $\Gamma_i$, both essential for the summation in (\ref{eq:E_c-distr}).
In contrast to earlier researchers who expanded (\ref{eq:E_c-distr}) by summing over selected final states with approximated matrix elements and incorporating multiple hole excitations, Brass {\it et al.} do not pursue the Fermi's golden rule approach and calculate instead linear response functions leveraging the Kubo formalism \cite{zagoskin_quantum_2014}. This entails the computation of Fourier-transformed causal Green's functions describing the time evolution of the quantum system during the EC process.

This \textit{ab initio} method is comprehensive, constructing multi-configurational electronic many-body wave functions for the Ho orbitals and one-particle wave functions for free electrons. It also includes a Hamiltonian that encodes the weak interaction and all Coulomb interactions between the nucleus and electrons, both bound and unbound. As a result, it inherently encompasses all contributions to the differential transition rate, which were previously selected and computed individually by other authors using various approximations.
While this method is undoubtedly a significant step forward in the understanding of the $^{163}$Ho EC process and of its calorimetric energy spectrum, it can unfortunately provide only mostly qualitative results. Excitation energies and broadenings are only approximate because of the limitations imposed by the adopted numerical methods.
Additionally, natural excitation widths are introduced phenomenologically, since the excitation lifetimes are determined also by the atomic de-excitations occurring \textit{after} EC and are thus not accounted for in the computed Green's functions.
Nevertheless, there are few important findings to be highlighted since they explain for the first time some features visible in experimental data: intra-orbital Coulomb scattering of electrons produces one- and two-hole excitations not present in previous works \cite{robertson_examination_2015,faessler_improved_2015,derujula_calorimetric_2016} and the line-widths are altered by the multiplet splitting of the resonances (as also explained in \cite{derujula_single_1983}).

One general conclusion of the works \cite{derujula_calorimetric_2016,brass_initio_2020} that most closely reproduce the experimental data is that the end-point region of the spectrum is free of peaks, and an analytical description requires only a manageable combination of smooth exponentials and the tails of Breit-Wigner resonances.

\section{Methods}
\label{sec:methods}

\subsection{Experimental set-up}
\label{ssec:setup}
The HOLMES experiment employs Transition-Edge Sensor (TES) microcalorimeters as detectors \cite{irwin_transition-edge_2005,ullom_review_2015}. These Mo/Cu bilayer TESs, with critical temperatures around $95\,$mK, are thermally coupled to (180$\times$180$\times 2)\,\mu$m$^3$ gold absorbers \cite{alpert_high-resolution_2019b}. The $^{163}$Ho isotope is embedded in the absorbers, positioned beside the sensors to avoid proximization effects. The detectors operate as calorimeters, where particle interactions in the absorber result in a measurable temperature rise.

The $^{163}$Ho isotope is produced through neutron irradiation of enriched $^{162}$Er\ samples, followed by chemical purification to eliminate contaminants \cite{heinitz_production_2018b}. Ion implantation embeds the isotope into the gold absorbers, enabling precise control over the implantation profile while minimizing contamination from unwanted isotopes.
The $^{163}$Ho implant is encapsulated between two 1\,$\mu$m-thick gold layers, which effectively\footnote{Absorption exceeds 99.99\% for the most energetic electrons, according to Geant4 simulations.} absorb the radiation emitted during the de-excitation of the resulting dysprosium atom.

HOLMES microcalorimeters are organized in 64-pixel arrays. Detector fabrication involves creating TES arrays, implanting $^{163}$Ho into the bottom gold layer of the absorber, depositing a top gold layer to encapsulate the isotope, and finally anisotropically etching the silicon substrate beneath each microcalorimeter to suspend it on a SiN membrane, which provides the thermal isolation required for operation.

The detector arrays are mounted in copper enclosures that house the bias circuitry and multiplexing chips, and are maintained at 40\,mK in a cryogen-free dilution refrigerator. TES signals are read out using microwave frequency-division multiplexing in the 4--8\,GHz band, employing flux-ramp linearized rfSQUIDs \cite{becker_working_2019b}.
A custom intermediate frequency (IF) board and an FPGA-based software-defined radio (SDR) system perform signal demodulation and data acquisition.

Energy calibration, when required, is performed using fluorescence X-rays generated by irradiating sodium chloride and aluminum targets with a $^{55}$Fe source. This provides calibration points at 1487\,eV (Al K$\alpha_1$ \cite{schweppe_accurate_1994}) and 2622\,eV (Cl K$\alpha_1$ \cite{deslattes_x-ray_2003}).

Typical detector performance includes rise times of $\sim$20\,$\mu$s, decay times of a few hundred $\mu$s, and energy resolutions of 5-10\,eV FWHM at 6\,keV, as determined by the heat capacity and noise characteristics \cite{bennett_impact_2025}.

\subsection{Data analysis}
\label{ssec:analysis}
The data analyzed in this work were collected during three separate measurement campaigns using one of the first ion-implanted HOLMES arrays.
This array was specifically prepared for high-statistics measurements of the $^{163}$Ho calorimetric spectrum and was ion-implanted to achieve a uniform distribution of activity across the pixels. The measured average activity was approximately 0.3\,Bq per detector, with a maximum of 0.6\,Bq.
The first campaign was dedicated to the calibration of the $^{163}$Ho spectrum, while the second campaign focused on a high-statistics measurement of the end-point region of the $^{163}$Ho spectrum to determine the neutrino mass. To ensure stable long-term data acquisition, an energy threshold of approximately 300\,eV was chosen for all detectors.
The third and final campaign targeted the lower energy range. Due to significantly higher noise rates at low energies, this measurement was performed over a few days using only a limited number of channels --specifically, those with the highest $^{163}$Ho activity and for which a threshold of approximately 30\,eV was achievable.

The raw spectra were processed through a series of steps \cite{borghesi_first_2022}, which include the rejection of spurious signals, amplitude estimation via optimal filtering, and correction for gain drifts by continuously monitoring the positions of calibration and $^{163}$Ho decay peaks over time (see \cite{alpert_most_2025,bennett_impact_2025} for further details).

In the first campaign, the array was initially measured using a fluorescence X-ray source to precisely determine the positions of the main peaks in the $^{163}$Ho spectrum. For this purpose, we analyzed data from the pixels with the highest $^{163}$Ho activity across multiple measurements, resulting in 51 datasets.
Energy calibration of their spectra was performed using the well-known energies of the K$\alpha_1$ X-rays of Al  and Cl, fitting a quadratic binomial to the calibration points.

\begin{figure*}[tbh]
  \centering
  \includegraphics[width=1.0\textwidth]{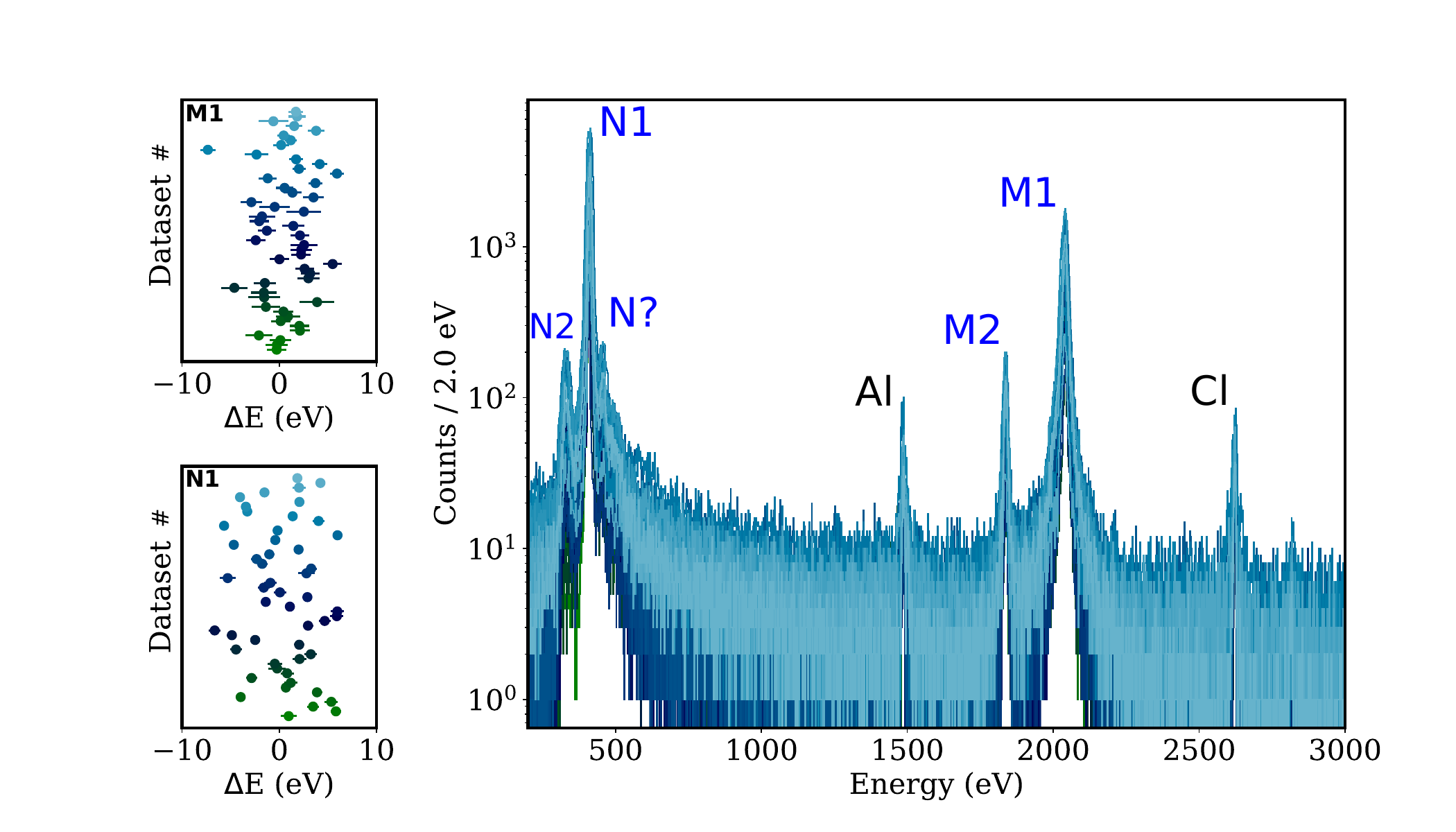}
  \caption{Superposition of the 51 calibrated spectra used to determine the positions of the main EC peaks (right). The peak labeled N? cannot be attributed to any single-hole excitations in Eq.\,\ref{eq:E_c-distr}. The spectra are calibrated using the energies of the Al and Cl K$\alpha$1 peaks, as indicated. The left panels show the distribution of the determined peak positions $E_0$ for the M1 (top) and N1 (bottom) lines across the 51 calibrated spectra. Error bars have been enlarged by a factor of 5 for clarity. The peak positions are plotted as $\Delta E = E_0 - 2040$\,eV and $\Delta E = E_0 - 411$\,eV for the M1 and N1 lines, respectively.}
  \label{fig:calibration}
\end{figure*}
Figure\,\ref{fig:calibration} shows the 51 calibrated spectra, where the peaks corresponding to EC from the M1, M2, and N1 shells are clearly visible. As illustrated in the insets of Figure\,\ref{fig:calibration}, the limited statistics of individual spectra and small deviations of the detector responses from a simple quadratic model result in a spread of the measured peak positions.
For each peak, we assume that the measured positions are distributed around the true value. To account for both the statistical uncertainties of individual measurements and the spread arising from detector non-linearity, we adopt an iterative procedure based on Bayesian learning \cite{lista_statistical_2016}.
A single calibrated spectrum is randomly selected to initiate the process. Bayesian parameter estimation is then performed using a Poisson model, with inference carried out via a Hamiltonian Markov Chain Monte Carlo algorithm implemented in STAN \cite{standevelopmentteam_stan_2024}. Given the limited statistics of individual spectra, each peak is modeled with an asymmetric Breit-Wigner function,
shifted by an offset $\delta_E$, and convolved with a Gaussian response function plus a flat background.
For the initial spectrum, the priors for all peak parameters (the energy position $E_0$, the linewidth $\Gamma$, and the asymmety parameter $\delta_{\mathrm{AS}}$) are modeled as broad Gaussian distributions. The remaining spectra are then fitted sequentially, using the posterior distributions from the previous fit as the priors for the three peak parameters. The priors for the other parameters --flat background, FWHM of the Gaussian response, normalization, and the offset $\delta_E$-- are not updated. The prior for $\delta_E$ is modeled as a normal distribution centered at zero, with a standard deviation determined from the observed spread in the data.

This procedure is repeated for each peak, and the mean and standard deviation obtained from the final spectrum are those reported in Table\,\ref{tab:peaks}, in agreement with the values measured in \cite{ranitzsch_first_2014}. These results are independent of the choice of the initial spectrum.
\begin{table}[h]
\centering
\caption{Peak parameters determined using the Bayesian learning procedure.
Only the peak positions $E_0$ are used to calibrate the partial spectra when no fluorescence X-ray source is available.
The peaks labeled N? and O? cannot be attributed to any single-hole excitations in Eq.\,\ref{eq:E_c-distr}; as discussed later in this article, they are associated with shake-off excitations (Table\,\ref{tab:e0data}).}
\label{tab:peaks}
\begin{tabular}{|c|c|c|c|}
\hline
Peak & Position $E_0$ [eV] & Width $\Gamma$ [eV] & Asymmetry $\delta_{AS}$ \\
\hline
\hline
M1 & $2040.8 \pm 0.3$ & $14.49 \pm 0.05$ & $1.306 \pm 0.006$ \\
M2 & $1836.4 \pm 0.8$ & $8.2 \pm 0.3$ & $1.03 \pm 0.05$ \\
N? & $454.5 \pm 0.1$ & $22.3 \pm 0.4$ & $0.62 \pm 0.02$ \\
N1 & $411.7 \pm 0.1$ & $5.57 \pm 0.03$ & $1.270 \pm 0.008$ \\
N2 & $329.0 \pm 0.1$ & $16.4 \pm 0.2$ & $0.69 \pm 0.01$ \\
O? & $61.0 \pm 0.8$ & $6.0 \pm 0.5$ & $1.000 \pm 0.009$ \\
O1 & $50.9 \pm 0.8$ & $2.4 \pm 0.4$ & $0.8 \pm 0.1$ \\
\hline
\end{tabular}
\end{table}

\begin{figure*} [htb]
\centering
  \includegraphics[width=0.9\textwidth]{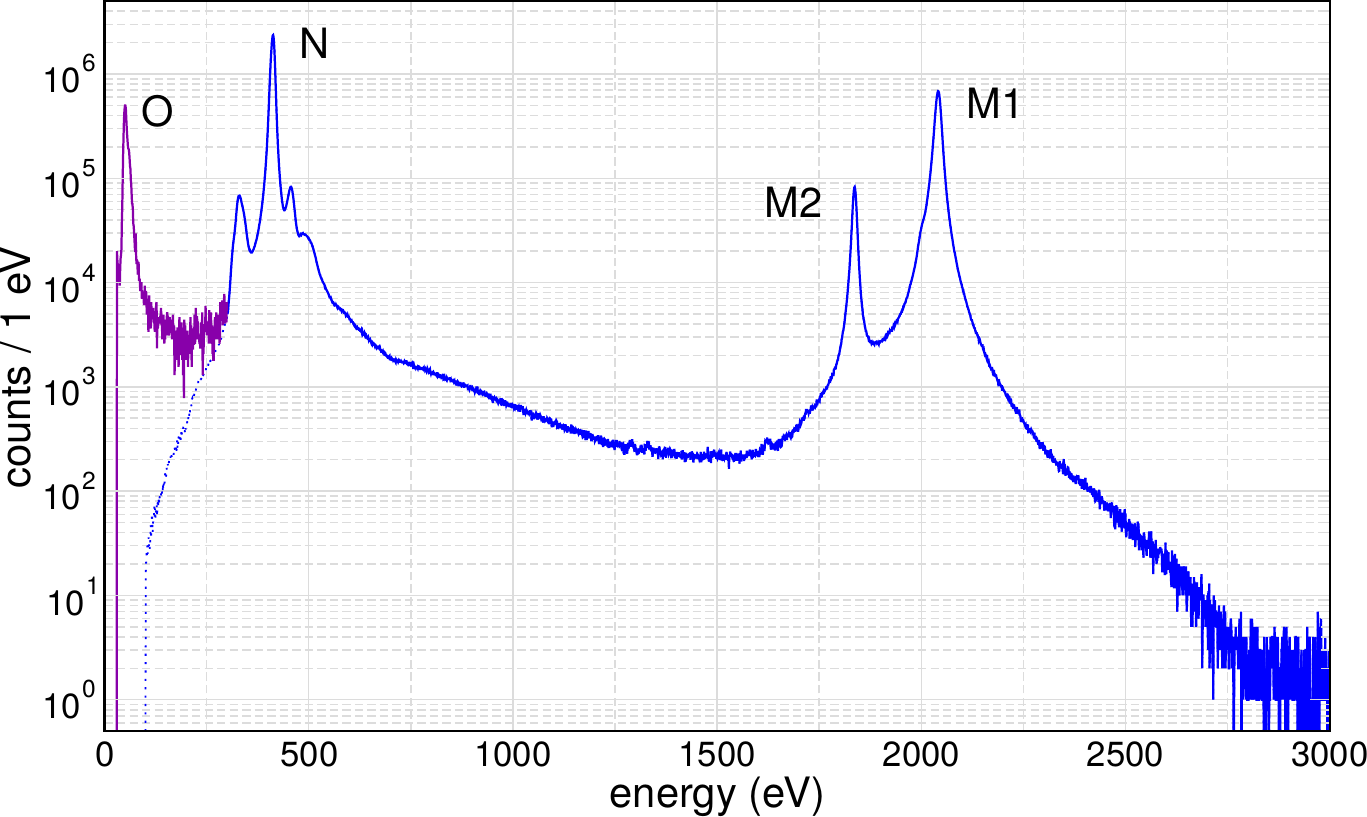}
\caption{Final calorimetric spectrum of the EC $^{163}$Ho decay. The blue histogram was measured in the neutrino mass campaign \cite{alpert_most_2025} and has an energy threshold at 300\,eV. The spectrum below 300\,eV is not used. The violet portion of the histogram below 300\,eV comes from the last campaign and has an energy threshold of about 30\,eV. The histogram has been scaled to precisely overlap to the blue one, using the ratio of the integrals above 300\,eV.}
\label{fig:spetot}
\end{figure*}

During the second campaign, data were collected to analyze the end-point in terms of neutrino mass (see \cite{alpert_most_2025} for details on the array and raw data processing). The resulting spectrum, shown as the blue curve in Fig.\,\ref{fig:spetot}, has an energy threshold of approximately 300\,eV and contains about $6\times10^{7}$\,events above this threshold. From this spectrum the endpoint is measured to be $E_0 = 2848^{+7}_{-6}$\,eV (see \cite{alpert_most_2025} for further details).

In the third campaign, measurements were performed on a subset of pixels with the lowest achievable thresholds, all below 30\,eV. The position of the O1 peak was determined using the same procedure as in the first campaign; however, the energy calibration was performed using a linear fit passing through the N1 peak and the origin. The resulting calibrated data, rescaled to ensure overlap with the spectrum from the second campaign above 300\,eV, are shown in Fig.\,\ref{fig:spetot} as the violet curve.

The more than 1000 individual spectra contributing to Fig.\,\ref{fig:spetot} exhibit varying Gaussian energy resolutions, with FWHM values ranging from 5\,eV to 10\,eV. Direct summation of these spectra can distort spectral features associated
with atomic de-excitations, particularly those with intrinsic widths comparable to or narrower than the resolution range. 
Fitting each individual spectrum to extract spectral features is impractical due to their small sample size and the similarity between intrinsic widths and resolution (see Table\,\ref{tab:peaks}).
A simultaneous fit of all spectra with common spectral components and varying Gaussian responses would be statistically unreliable and computationally prohibitive, especially since the spectral components are initially largely unknown and only roughly approximated by the simplified model used to obtain the parameters in Table\,\ref{tab:peaks}.
To address this, we implemented a more robust and computationally efficient iterative unfolding method to correct for the Gaussian response of each individual spectrum prior to summation.
The unfolding procedure is based on Bayes’ theorem, following the approach described in \cite{dagostini_multidimensional_1995}, and is implemented using the PyUnfold framework \cite{bourbeau_pyunfold_2018}.
The observed distribution of event energies (effects) arises from the smearing of the true energy distribution (causes) due to the detector response. The goal of the unfolding process is to estimate the true distribution of causes ($C_i$), starting from the observed data, which consists of the measured effect frequencies $n(E) = \{n(E_1), n(E_2), . . . , n(E_{N_E})\}$, with a total number of observed events $N_{obs}$, and a total number of bins $N_E$.
Assuming a prior energy distribution $P(C_i)$, the unfolding matrix is constructed using Bayes' theorem:
\begin{equation} \label{eq:aa}
P\left(C_i \mid E_j\right)=\frac{P\left(E_j \mid C_i\right) \cdot P\left(C_i \right)}{\sum_k P\left(E_j \mid C_k\right) \cdot P\left(C_k \right)}
\end{equation}
which gives the probability that a measured energy $E_j$ originated from a true energy $C_i$.
Here, $P(E_j \mid C_i)$ is the smearing matrix, estimated via Monte Carlo simulations based on the measured detector response and the model of Eq.\,\ref{eq:E_c-distr}.
The unfolded energy distribution is given by convolving the unfolding matrix with the reconstructed energy distribution via:
\begin{equation} \label{eq:bb}
\hat{n}(C_i) = \frac{1}{\epsilon_i}\sum_j n(E_j) P\left(C_i \mid E_j\right)
\end{equation}
where $\epsilon_i=\sum_{j} P(E_j | C_i)$ is the efficiency for detecting the cause $C_i$ in any effect bin.
The analysis is performed iteratively: starting with an initial uniform prior $P(C_i)$ and assuming 100\% efficiency in each bin, the unfolding matrix is computed according to Eq.\,(\ref{eq:aa}), yielding a posterior distribution using Eq.\,(\ref{eq:bb}). At each iteration, the estimate of the true distribution and efficiencies are progressively refined:
\begin{equation}
\begin{aligned}
& \hat{P}\left(\mathrm{C}_i\right) \equiv P\left(\mathrm{C}_i \mid n(\mathrm{E})\right)=\frac{\hat{n}\left(\mathrm{C}_i\right)}{\hat{N}_{\text {true }}} \\
& \hat{\boldsymbol{\epsilon}}=\frac{N_{\text {obs }}}{\hat{N}_{\text {true }}}
\end{aligned}
\end{equation}
where $\hat{N}_{true} = \sum_i \hat{n}(C_i)$ is the best estimate of the true total number of events, $\hat{P}(C_i)$ is the updated probability of causes and $\hat{\epsilon}$ is the updated overall efficiency.
This updated distribution is used as the subsequent prior estimate, a procedure that ends once variations on $\hat{n}\left(\mathrm{C}_i\right)$ from one iteration to the next are negligible.
The difference between successive iterations was quantified using the Kolmogorov–Smirnov test. Convergence was defined as being reached when the test statistic yielded a $p$-value smaller than 0.001. This threshold was determined through dedicated Monte Carlo simulations to ensure optimal convergence behavior.
The procedure is applied to each of the $\sim$1000 individual spectra, using smearing matrices tailored to their respective energy resolutions. Once the unfolding is performed on all detectors, the mean value for each bin is computed across all normalized spectra, producing the final unfolded spectrum. The statistical uncertainty for each bin in the combined unfolded spectrum is estimated as the standard deviation of the mean of the corresponding bin values from the individual unfolded spectra. The systematic uncertainty accounts for both the uncertainty in the individual spectrum energy resolution and the choice of the $p$-value threshold for terminating the iterations. Specifically, the systematic uncertainty was estimated as the difference between the unfolded spectra obtained by varying the energy resolution by \(\pm0.25\)\,eV, thus accounting for both the uncertainty in its determination and any possible weak energy dependence, and the stopping $p$-value threshold by \(\pm0.0001\). These differences were then combined in quadrature with the statistical uncertainty.
Here, we do not include the systematic uncertainty on the energy scale of the individual spectra, which is discussed in detail in \cite{alpert_most_2025} for the end-point region. In the region between the N and M peaks, this uncertainty is estimated to be negligible. The appearance in the final spectrum shown in Fig.\,\ref{fig:spetot} of four faint resonances between 1250\,eV and 1750\,eV (see Table\,\ref{tab:fit_params_asym_bw}, components 13--16) provides indirect evidence that the uncertainty on the energy scale of the individual spectra is negligible.
The unfolding procedure was applied to the data shown in Fig.\,\ref{fig:spetot}, focusing on the energy range 30--2900\,eV. 
The resulting unfolded spectrum, normalized to unit area, forms the basis for fitting the individual spectral components.

\section{Results}
\subsection{Models}
\label{sec:model}
The calorimetric spectrum of the $^{163}$Ho EC decay is expected to consist of a series of discrete lines and continuous components.

The discrete peaks arise from excitations that de-excite with radiation emission into the continuum only after the initial capture interaction.
Starting excitations can involve one or more holes, contributing to the spectrum with a peak at an energy corresponding to the sum of the binding energies of the electrons that left the holes. Comparison of experimental spectra with theoretical predictions reveals that there is currently no reliable method to predict the probabilities of the various possible starting excitations.
Moreover, {\it ab initio} studies \cite{brass_initio_2018,brass_initio_2020} indicate that the energy of these excitations is challenging to calculate accurately. Intrashell Coulomb interactions
shift the excitation energies, and computational methods lack the precision needed for reliable predictions. The expected result is a complex multiline structure (see Fig.\,5 in \cite{brass_initio_2020}),
 which can be smoothed by the detector energy resolution, the intrinsic line shape -- still not fully determined by {\it ab initio} calculations -- and statistical uncertainties.
A direct comparison between the phenomenological spectrum obtained from our fit and the {\it ab initio} calculation is presented later in Sec.\,\ref{sec:discussion} (see Fig.\,\ref{fig:cfr}).

Regarding the shake-off continuous contributions, {\it ab initio} calculations and the results in \cite{derujula_calorimetric_2016} highlight their importance in understanding the main peak tails and the continuum far from the peaks (see discussion in Sec.\,\ref{sec:discussion} and Fig.\,\ref{fig:brass}). However, they do not provide analytical or parametric expressions for the shake-off components of the calorimetric spectrum.

Taken together, even the most accurate {\it ab initio} calculations fail to reproduce all the features of the experimental $^{163}$Ho EC calorimetric spectrum, making them unsuitable for directly fitting the experimental data; a comparison with our phenomenological fit is shown later in Sec.\,\ref{sec:discussion} (Fig.\,\ref{fig:cfr}).
Despite these limitations, these calculations offer valuable insights into the expected contributions, especially their positions and intensities. Leveraging this understanding, we fit the HOLMES high-statistics energy spectrum using phenomenological analytic expressions to model all identified single and double atomic excitations.
Below, we describe the functions employed to fit the various components: symmetric and asymmetric resonance peaks, as well as shake-off continuous distributions.

Peaks are, in the first approximation, expected to have a Breit-Wigner shape, whose width is related to the lifetime of the excitation. However, as shown by theoretical work \cite{brass_initio_2020,derujula_calorimetric_2016}, the situation is more complex. In addition to uncertainties in position and intensity, peaks may exhibit strong deviations from the Breit-Wigner shape due to quantum mechanical interferences between resonances, threshold effects, and other phenomena \cite{fano_effects_1961,riisager_low-energy_1988,derujula_calorimetric_1982,derujula_calorimetric_2016,derujula_single_1983,brass_initio_2020,brass_initio_2021a}.
For these reasons, we adopted the following phenomenological parametrization for the intrinsic shape of peaks:
\begin{align}
\mathrm{BW}(E| E_0, \Gamma, \delta_{AS},p,E_{th}) & = \frac{\delta_{AS}}{\pi(1+\delta_{AS})}
\frac{\Gamma _{L}}{(E - E_{0})^{2} + \Gamma _{L}^{2}/4} \left(\frac{E - E_{th}}{E_0 - E_{th}} \right)^p \nonumber \\
\mbox{ for } E_{th} \leq E \leq E_{0}, \label{eqn:bw} \\
\mathrm{BW}(E| E_0, \Gamma, \delta_{AS},p,E_{th}) & =  \frac{1}{\pi(1+\delta_{AS})}\frac{\Gamma
  _{R}}{(E - E_{0})^{2} + \Gamma _{R}^{2}/4}, \nonumber \\
\mbox{ for } E > E_{0}, \label{eqn:bw_right} \\
\Gamma _{R} & =  \frac{2 \Gamma}{1 + \delta _{AS}}, \: \Gamma _{L} = 2 \Gamma - \Gamma _{R}. \nonumber
\end{align}
This parametrization accounts for the strongly asymmetric shape of the main peaks. In particular, the last term in Eq.~(\ref{eqn:bw}) was introduced in \cite{riisager_low-energy_1988} to model the atomic phase space in the de-excitation process, effectively reproducing the observed strong suppression of the left tail of the Breit-Wigner shape for certain peaks. 
Here, $E_{th}$ denotes the threshold energy below which the left-tail suppression occurs, and $p$ is a shape parameter controlling the steepness of this suppression; their detailed definition can be found in \cite{riisager_low-energy_1988}.
Simple asymmetric and fully symmetric shapes are recovered for $p=0$ or $p=0$ and $\delta_{AS}=1$, respectively\footnote{\label{bw_norm}Eq.\,(\ref{eqn:bw}) is properly normalized to unity only if $p=0$, in all other cases the amplitude of the component will not be equal to its integral in the spectrum.}.

Previous works \cite{derujula_calorimetric_2016,faessler_neutrino_2017} calculated the spectral shape of the electron emitted during the shake-off process in $^{163}$Ho by building on the approach of Intemann and Pollock \cite{intemann_electron_1967}, which involves evaluating integrals of electron wave functions. In contrast, we adopted the approach used in \cite{robertson_shakeup_2020} to describe the shake-off satellites observed in the internal conversion electron spectrum of $^{83}$Kr.
This approach utilizes the analytic expressions for the shake-off spectral distributions derived by Levinger \cite{levinger_effects_1953} for large-Z hydrogenic atoms. Specifically, we followed \cite{robertson_shakeup_2020} by employing the shake-off shape calculated for $1s$ holes for any $nl$ excitation in Dy and leveraging their analytical approximation of the convolution of the shake-off function with the Breit-Wigner width of the two-hole state.
The parametrization we used for the shake-off distribution is as follows:
\begin{equation}
\label{eq:sof}
\mathrm{SOF}(E|E_0,\Gamma,E_b) = C \left[\frac{1}{\pi} \arctan\left(2 \frac{E - E_0}{\Gamma}\right) + \frac{1}{2}\right] P_{1s}\left(\kappa \right),
\end{equation}
where 
$E_0 = B[\mathrm{H1}] + B[\mathrm{H2}]$ is the sum of the binding energies of the two holes involved, $W = E - E_0$ is the kinetic energy of the outgoing shake-off electron, $\kappa = \sqrt{E_b/|W|}$, and $E_b$ denotes the nominal binding energy of the shake-off electron.
\begin{equation}
  P_{1s}\left(\kappa\right) = \frac{\kappa^8}{(\kappa^2 + 1)^4} \cdot e^{-4\kappa \arctan\left(\frac{1}{\kappa}\right)} \cdot \frac{1}{1 - e^{-2\pi \kappa}}.
\end{equation}
Given that $E_b$ is treated as a fit parameter, in general it loses its physical meaning of binding energy. 
Since an analytic normalization of Eq.~\eqref{eq:sof} is not possible, we set $C=1$, as the integral of Eq.~\eqref{eq:sof} is always of order unity.

However, due to the approximations described above, $E_b$ no longer retains its original interpretation as the binding energy; instead, it describes the energy scale of the decay in the electron energy distribution. We opted for the above parametrization despite the fact that a simple double exponential function could also adequately describe the data \cite{alpert_most_2025}. The $\Gamma$ and $E_b$ parameters in Eq.~\eqref{eq:sof} effectively act as the two exponential constants, representing the rising and decaying scales of the distribution, respectively.

The total calorimetric spectrum in Eq.~(\ref{eq:E_c-distr}) for $m_{\nu}=0$\,eV/$c^2$ can now be expressed as a linear combination of Breit-Wigner resonances (as defined in Eqs.\,\ref{eqn:bw} and \ref{eqn:bw_right}) and shake-off spectra (from Eq.\,\ref{eq:sof}), each multiplied by the phase space factor:
\begin{equation}
\label{eq:tot}
\begin{aligned}
N(E_c) = & \left[\sum_i A^{BW}_i \mathrm{BW}(E_c, E_{0_i}, \Gamma_i, \delta_{{AS}_i}, p_i, E_{{th}_i}) \right. \\
& \left. + \sum_i A^{SOF}_i \mathrm{SOF}(E_c, E_{0_i}, \Gamma_i, E_{b_i}) \right] \left( Q - E_c\right)^2,
\end{aligned}
\end{equation}

\subsection{Spectrum fit}
\label{ssec:fit}
\begin{figure*} [tb!]
  \centering
  \includegraphics[width=0.99\textwidth]{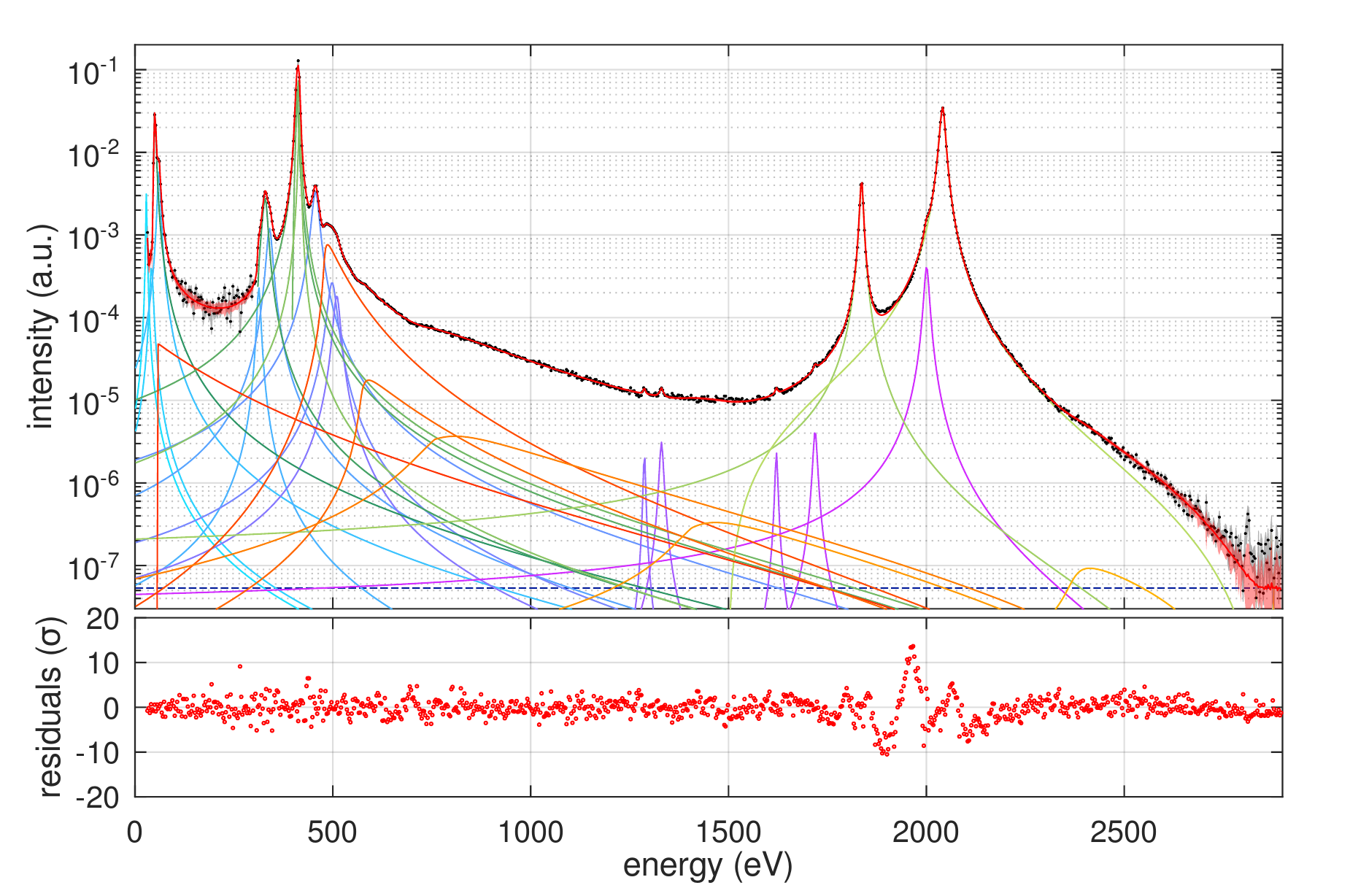}
\caption{Unfolded spectrum (black dots) with the fitted model (red curve) after the STAN fit. Uncertainties on the unfolded spectrum are shown as a gray band, which is barely visible due to the small size of the statistical and systematic errors. The band is primarily visible  below 300\,eV and near the endpoint, where the statistical fluctuations in the data are larger. The red line and reddish band represent the mean and standard deviation of the STAN-generated model, based on the posterior distributions. Coloured curves indicate the 25 fitted components: symmetric Breit-Wigner resonances (violet), asymmetric Breit-Wigner resonances (green), and shake-off spectra (red). The dashed dark blue line denotes the constant background term.}
\label{fig:full_bare}
\end{figure*}

\begin{figure*}[hbt!]
  \centering
  \includegraphics[width=0.99\textwidth]{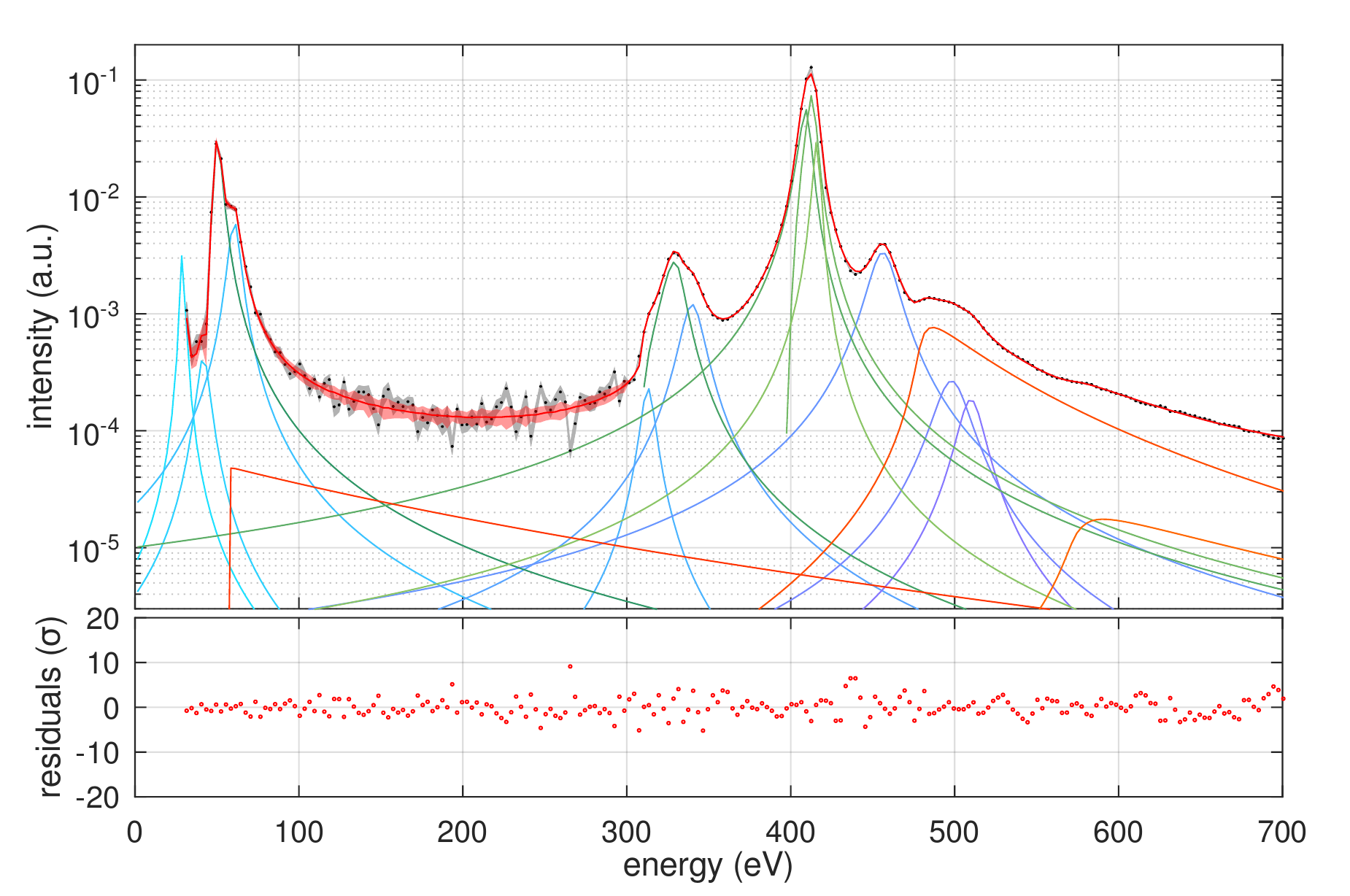}
\caption{Zoomed view of the O and N groups in the unfolded spectrum, overlaid with the fitted model. The color coding matches that of Fig.\,\ref{fig:full_bare}.}
\label{fig:NO_bare}
\end{figure*}
The fitting procedure was performed in two steps. First, the spectral components were introduced incrementally using a trial-and-error approach, with their parameters adjusted through controlled Minuit2 optimization \cite{hatlo_developments_2004}.
In this step, to account for the phase space factor, the unfolded spectrum (consisting of 956 bins) was divided by $(Q - E)^2$, where $Q$ is the endpoint energy of the $^{163}$Ho EC decay. This transformation flattens the spectrum, particularly at higher energies, and facilitates the fitting process.
The procedure began with the lower energy group of the O lines, progressively adding the N and M groups while treating the lower energy groups as a fixed background. Finally, the components between the N and M groups, primarily the shake-off contributions, were adjusted. This process was iterated several times until a satisfactory fit was achieved with 19 Breit-Wigner peaks and 6 shake-off spectra.
An additional Breit-Wigner peak was included below the experimental threshold at a fixed position $E_0 = 29$\,eV to account for the expected O2 capture peak contribution to the background.\footnote{\label{foot:O2}Although the O2 peak is typically reported in the literature at about 26\,eV, we chose a slightly higher position for this sub-threshold peak, as this value is consistent with the position observed in the high-statistics ECHo measurement\cite{echo-com,nucciotti_present_2024}. We note that the choice of its position affects only its intensity (see Table\,\ref{tab:fit_params_asym_bw}).} Its amplitude and width $\Gamma$ were optimized to ensure a smooth connection with the experimental data at threshold.
This peak was subsequently treated as a fixed background component in the next fitting step.

The second step consisted of a simultaneous Bayesian estimation of all parameters (25 spectral components with 103 free parameters, plus a constant term accounting for background unrelated to the $^{163}$Ho decay \cite{borghesi_background_2024}\footnote{The background described in \cite{borghesi_background_2024} is approximately constant only above 2000\,eV; from threshold to 2000\,eV, it decreases by about a factor of 10. However, except near the endpoint where the constant approximation is fully justified, the background remains negligible.}), using a Normal likelihood with a width given by the combined statistical and systematic uncertainties from the unfolding procedure.
Normally distributed priors were assigned to each parameter, centered on the values obtained in the previous fitting step, with standard deviations set to the ones reported by Minuit2.
The posterior distribution was explored using a Hamiltonian Markov Chain Monte Carlo algorithm implemented in STAN \cite{standevelopmentteam_stan_2024}. In this step, the unfolded spectrum, including the phase space factor, was fitted.

When fitting the unfolded spectrum, a background due to pile-up was neglected because of the very low $^{163}$Ho activity in the detectors.
In the data shown in Fig.\,\ref{fig:spetot},
the intensity of the pile‑up spectrum is expected to be at least five orders of magnitude lower than that of the single‑event spectrum \cite{alpert_most_2025}.

The final result for the unfolded spectrum,
is shown in Fig.\,\ref{fig:full_bare}.
The parameters estimated through this procedure are listed in the Appendix\,\ref{app:tables}, while Appendix\,\ref{app:corr} shows their correlations.
A close inspection of the residuals in the M region suggests that the phenomenological model may be incomplete. The observed deviations could be attributed to limitations in the asymmetric Breit-Wigner peak parametrization of Eq.\,(\ref{eqn:bw}), which may require further refinement or a modified analytical expression.

Fig.\,\ref{fig:NO_bare} presents an enlarged view of the O and N groups, highlighting their considerable complexity.
Despite efforts to include components guided by theoretical models of $^{163}$Ho EC, the phenomenological approach may not fully capture all underlying physical processes. In particular, shake-off components could potentially mimic the features of strongly asymmetric Breit-Wigner peaks, leading to possible ambiguities or degeneracies in their interpretation (see for example component 8 at 412.9\,eV in Table\,\ref{tab:fit_params_asym_bw}).
A comparison of the results in Tables\,\ref{tab:peaks} and \ref{tab:fit_params_asym_bw} highlights how the more detailed modeling of the Breit-Wigner peaks in the M and N groups, together with the increased statistics, leads to significant changes in the peak parameters. It is important to emphasize the different objectives of the two analyses: the fit reported in Table\,\ref{tab:peaks} was primarily intended to provide accurate peak positions for calibration purposes, employing a simplified functional form for the peaks that is fully adequate for the partial spectra with their limited statistics (approximately 1/1000 of the final spectrum).

\section{Discussion}
\label{sec:discussion}
\begin{figure*}[t]
  \includegraphics[width=0.95\textwidth]{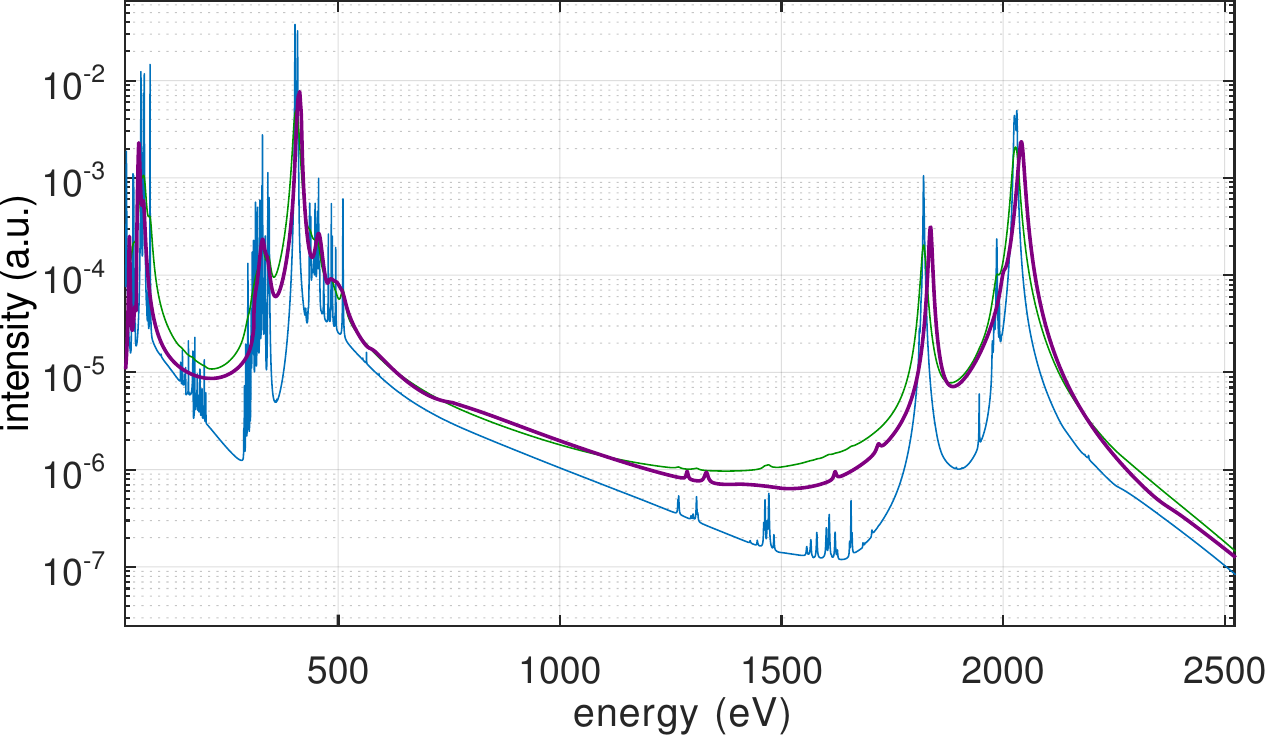}
  \caption{Comparison between the phenomenological spectrum obtained from the fit with STAN (purple) and the spectrum derived from {\it ab initio} calculations \cite{brass_initio_2020} (blue), with data kindly provided by M.\,Brass. To facilitate the comparison, the green curve is obtained by applying a uniform Lorentzian broadening with $\Gamma = 10$\,eV to the blue curve.}
\label{fig:brass}
\label{fig:cfr}
\end{figure*}
Fig.\,\ref{fig:cfr} presents the phenomenological model developed in this work, alongside the results of the {\it ab initio} calculations from \cite{brass_initio_2020} (see the caption for further details). The comparison demonstrates that, while the {\it ab initio} calculation closely reproduces the real EC spectrum, it cannot be used as-is for the analysis of experimental data or for neutrino mass sensitivity studies.
The first observation is that the {\it ab initio} calculations (shown in green) do not fully capture the natural broadening of the resonances.
To facilitate comparison, the blue curve is obtained by convolving the theoretical spectrum with a Breit-Wigner function of 10\,eV width.
Furthermore, the resonance energies in the theoretical spectrum do not align with those observed in the experimental data, particularly for the main single-hole resonances.
Both discrepancies can be attributed to limitations and approximations inherent in the {\it ab initio} approach, as discussed in \cite{brass_initio_2020}.

Despite the aforementioned limitations, the comparison in Fig.\,\ref{fig:cfr}, supported by \cite{brass_initio_2018,brass_initio_2020}, remains highly valuable for guiding the interpretation of the components observed in the spectrum.
Theoretical predictions reported in \cite{robertson_examination_2015,faessler_neutrino_2017} do not fully reproduce the features observed in the experimental data, whereas those in \cite{derujula_calorimetric_2016} show some improvements; however, in that work, the intensities were empirically adjusted to achieve better agreement with the data.
\begin{table}[htbp]
    \centering
    \caption{Interpretation of the components identified through the fit of the unfolded spectrum in terms of single and double excitations. $E_0$ denotes the positions found by the fit (see Tables\,\ref{tab:fit_params_asym_bw} and \ref{tab:fit_params_shakeoff}); the value in the first row has no error as it is not obtained from the fit. The second column indicates the component type: BWS and BWA refer to symmetric and asymmetric Breit-Wigner resonances, respectively, while SOF denotes shake-off spectra. H1 and H2 correspond to the Dy excitations following EC, with their binding energies listed as $B[H1]$ and $B[H2]$, respectively. The total excitation energy is given by $B[H1]+B[H2]$. The column ``Dev.'' lists the deviations between the calculated excitation energy and the fitted $E_0$ value. The last column reports the component intensity, obtained from the numerical integral as explained in the text, and normalized to the M1 intensity.}
    \label{tab:e0data}
    \begin{tabular}{|l|c|c|c|c|c|c|c|c|c|}
    \hline
    $E_0$ & Type & H1 & H2 & B[H1] & B[H2] & B[H1]+B[H2] & Dev. & Int. Rel. \\
    (eV) &  &  &  & (eV)& (eV) & (eV) & (eV) &  \\
    \hline\hline
29 & BWS & O2 & & 26.3 & & 26.3 & 2.7 & 0.0015966(5) \\
\hline
41.7(8) & BWS & O2 & N6 & 26.3 & 8.6 & 34.9 & 6.8 & 0.00057(5) \\
50.05(4) & BWA & O1 & & 49.9 & & 49.9 & 0.1 & 0.0269(3) \\
58.45(6) & SOF & O1 & N6 & 49.9 & 8.6 & 58.5 & -0.1 & 0.0038(8) \\
60.54(6) & BWS & O1 & N6 & 49.9 & 8.6 & 58.5 & 2.0 & 0.0079(1) \\
312.9(2) & BWS & N4 & N5 & 153.6 & 160 & 313.6 & -0.7 & 0.00043(3) \\
329.22(8) & BWA & N2 & & 333.5 & & 333.5 & -4.3 & 0.0072(1) \\
339.93(9) & BWS & N2 & N6 & 333.5 & 8.6 & 342.1 & -2.2 & 0.00374(9) \\
409.65(8) & BWA & N1 & & 414.2 & & 414.2 & -4.5 & 0.0984(7) \\
412.9(1) & BWA & N1 & N6 & 414.2 & 8.6 & 422.8 & -9.9 & 0.090(2) \\
416.4(2) & BWA & N1 & N7 & 414.2 & 5.2 & 419.4 & -3.0 & 0.033(1) \\
456.15(3) & BWS & N3 & N5 & 293.2 & 160 & 453.2 & 3.0 & 0.01431(4) \\
479.6(1) & SOF & N2 & N4 & 333.5 & 160 & 493.5 & -13.9 & 0.0087(2) \\
498.1(4) & BWS & N2 & N4 & 333.5 & 160 & 493.5 & 4.6 & 0.0014(1) \\
509.9(3) & BWS & N2 & N4 & 333.5 & 160 & 493.5 & 16.4 & 0.00076(9) \\
573.4(7) & SOF & N1 & N5 & 414.2 & 160 & 574.2 & -0.8 & 0.0019(2) \\
757(3) & SOF & N1 & N2 & 414.2 & 343.5 & 757.7 & -0.5 & 0.0044(2) \\
1287.3(7) & BWS & M5 & & 1292 & & 1292.0 & -4.7 & 0.000007(2) \\
1330.6(5) & BWS & M4 & & 1333 & & 1333.0 & -2.4 & 0.000018(2) \\
1406(6) & SOF & M5 & N4 & 1333 & 49.3 & 1382.3 & 23.3 & 0.0015(2) \\
1621.0(7) & BWS & M5 & N2 & 1292 & 343.5 & 1635.5 & -14.5 & 0.000014(2) \\
1718.8(5) & BWS & M5 & N1 & 1292 & 432.4 & 1724.4 & -5.6 & 0.000044(4) \\
1863.25(3) & BWA & M2 & & 1842 & & 1842.0 & -5.7 & 0.05189(4) \\
2000.66(7) & BWS & M2 & N4 & 1842 & 160 & 2002.0 & -1.3 & 0.00957(6) \\
2041.25(2) & BWA & M1 & & 2047 & & 2047.0 & -5.8 & 1.0000(4) \\
2371(2) & SOF & M1 & N2 & 2047 & 343.5 & 2390.5 & -19.1 & -- \\
\hline
\end{tabular}
\end{table}
Starting from the results listed in the tables in the appendices an attempt was made to provide an interpretation of all identified components in terms of single- or double-hole excitations.
Table\,\ref{tab:e0data} reports, for each fitted component, the excitation energy $E_0$ obtained from the fit alongside the corresponding calculated value for the assigned excitation.
There is no consensus in the literature regarding the calculation of excitation energies. In this work, we adopt the approach of \cite{robertson_examination_2015}, assigning the binding energy of Dy to the innermost shell (H1) and that of Ho to the less-bound shell (H2). This method yields the minimum root mean square deviation. However, we still observe deviations of several tens of electron volts in some cases; these may reflect limitations in the theoretical calculations, or that our choice of component type or the assignment of shells for the transition is not correct.

\begin{figure*}[tbh]
  \centering
  \includegraphics[width=0.45\textwidth]{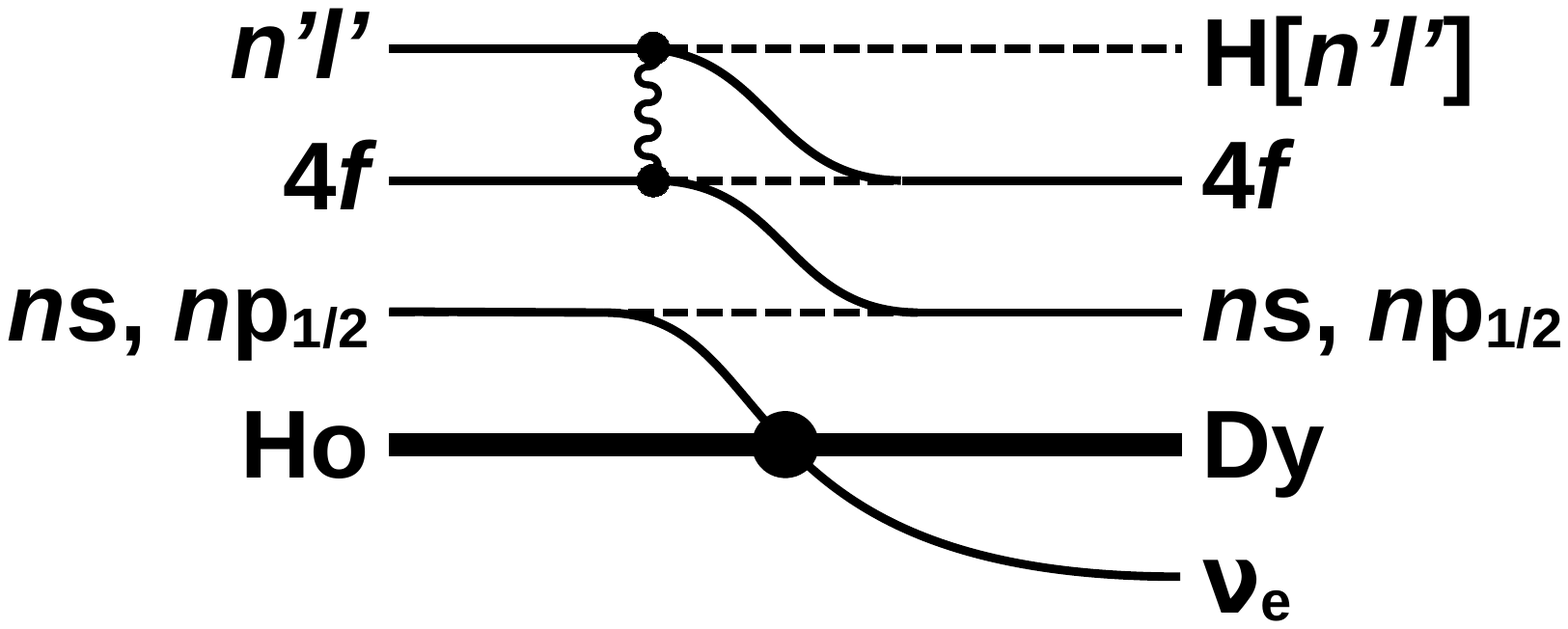}
   \includegraphics[width=0.45\textwidth]{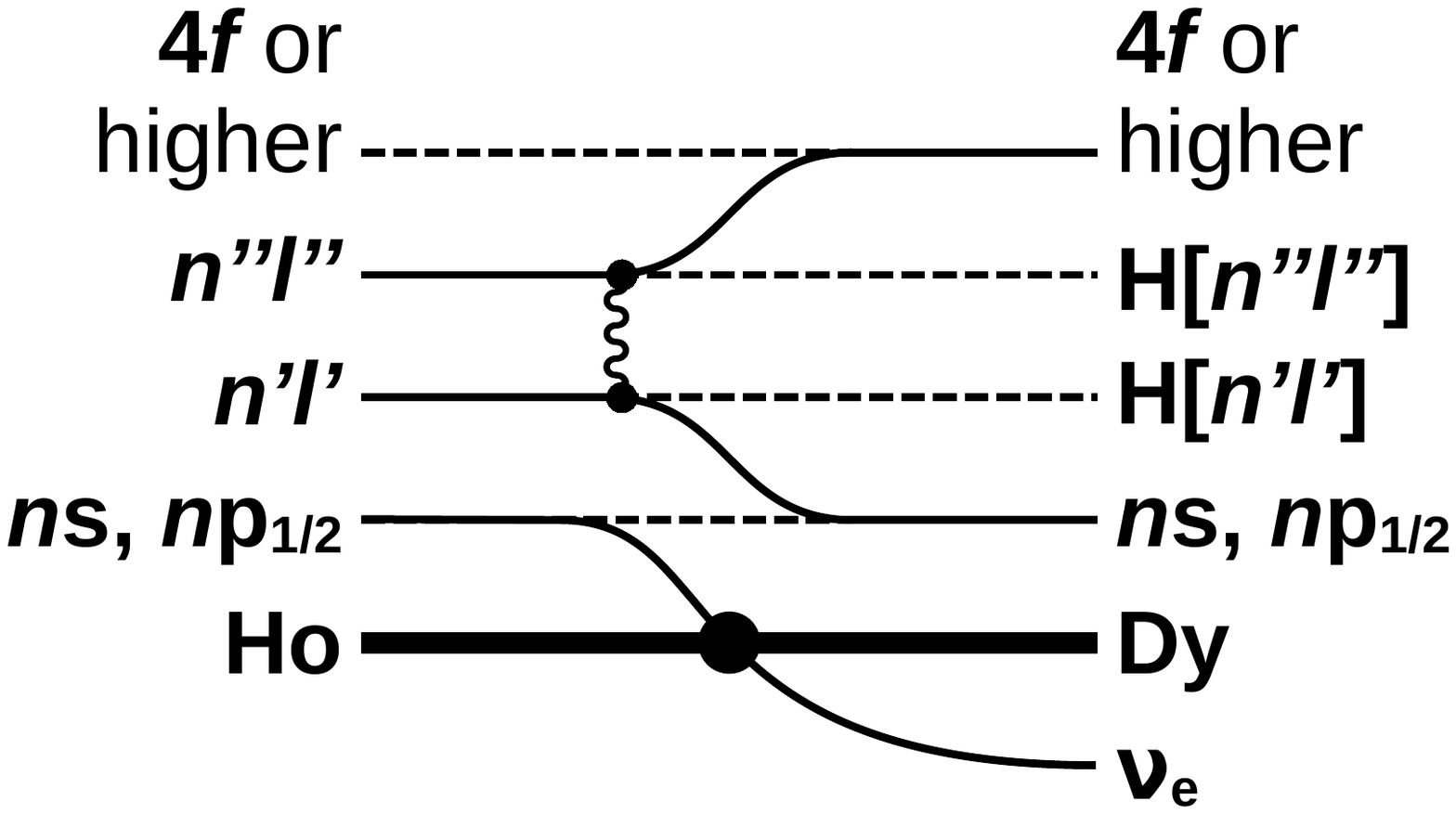}
\caption{Schematic representation of atomic excitations leading to holes in shells not directly accessible via EC due to Coulomb repulsion: single-hole creation (left), and double-hole creation via shake-up (right). Double-hole states can also arise from shake-off processes, analogous to shake-up, but with one electron scattered to an unbound state (not shown). Here, $n=4,5,6$ denote the principal quantum numbers of the core holes. For completeness, we note that shake-up transitions can be further classified as Coster-Kronig ($n=n^\prime$) or super Coster-Kronig ($n=n^\prime=n^{\prime\prime}$) \cite{derujula_two_2013}.}

\label{fig:coulomb}
\end{figure*}

Notably, as predicted in \cite{brass_initio_2018,brass_initio_2020}, the model components listed in Table\,\ref{tab:e0data} include excitations with the main hole H1 in shells that are not expected to be directly accessible via EC according to the single-hole model \cite{derujula_calorimetric_1982}, or even models incorporating higher-order excitations \cite{robertson_shakeup_2020,faessler_neutrino_2017,derujula_calorimetric_2016}, such as the M3, M4, M5, N3, N4, and N5 shells.
These additional features correspond to inter-core relaxation processes, as anticipated in \cite{brass_initio_2018}, where Coulomb repulsion transfers angular momentum from the core hole to a $4f$ electron in the valence shell.
Coulomb inter-core interactions play a role in both single-hole excitations and shake-off or shake-up processes, leading to core holes (H1) in shells not directly accessible via EC \cite{brass_initio_2018,brass_initio_2020} (see Fig.\,\ref{fig:coulomb}).
Unfortunately, due to the still limited available statistics, particularly below 300\,eV, only the M4 and M5 single-hole excitations are clearly visible.

{\it Ab initio} calculations also indicate that Coulomb interactions lead to significant multiplet splitting of all components.
Notably, the {\it ab initio} calculations reproduce the substantial intensity observed in the group of excitations to the right of the N1 resonance and on the left flank of the M1 resonance. These two groups arise from shake-up processes involving core holes in N2 and N3 or M2, respectively, and are not predicted with comparable intensity by other authors.
As previously noted for the O and N groups, the structure is quite complex in both the experimental data and the {\it ab initio} calculations.
In particular, the latter predicts several multiplets that are not resolved in our experimental data, making direct comparison and component interpretation challenging. Our spectrum appears to be interpretable with a limited number of dominant structures, either due to limited statistics or to the natural linewidths.
In Table\,\ref{tab:e0data}, we attempt to assign double-hole excitations to these few identified resonances, keeping in mind that the actual fine structure may not be fully resolved. For example, two resonances are associated with the same N2N4 shake-up, implicitly acknowledging the multiplet nature of these excitations.

A fine structure is also predicted for the single-hole resonances, which could provide an additional explanation for the relatively poor fit of the M1 peak.

Finally, Fig.\,\ref{fig:cfr} illustrates that the {\it ab initio} calculations, once unbound states are included in the model as in \cite{brass_initio_2020}, also account for the asymmetry of the single-hole resonances and the enhanced intensity observed on their high-energy tails.
However, \cite{brass_initio_2020} does not provide a sufficiently detailed analysis of the single shake-off processes required for a thorough comparison with the corresponding components listed in Table\,\ref{tab:e0data}.
In our analysis, we observe that not every two-hole shake-up excitation is accompanied by the corresponding shake-off excitation, as suggested in \cite{derujula_calorimetric_2016}. In fact, we identified only six shake-off components, a finding that appears to be supported by comparison with the {\it ab initio} spectrum.

\begin{figure*}[hbt]
\centering
  \includegraphics[width=0.85\textwidth]{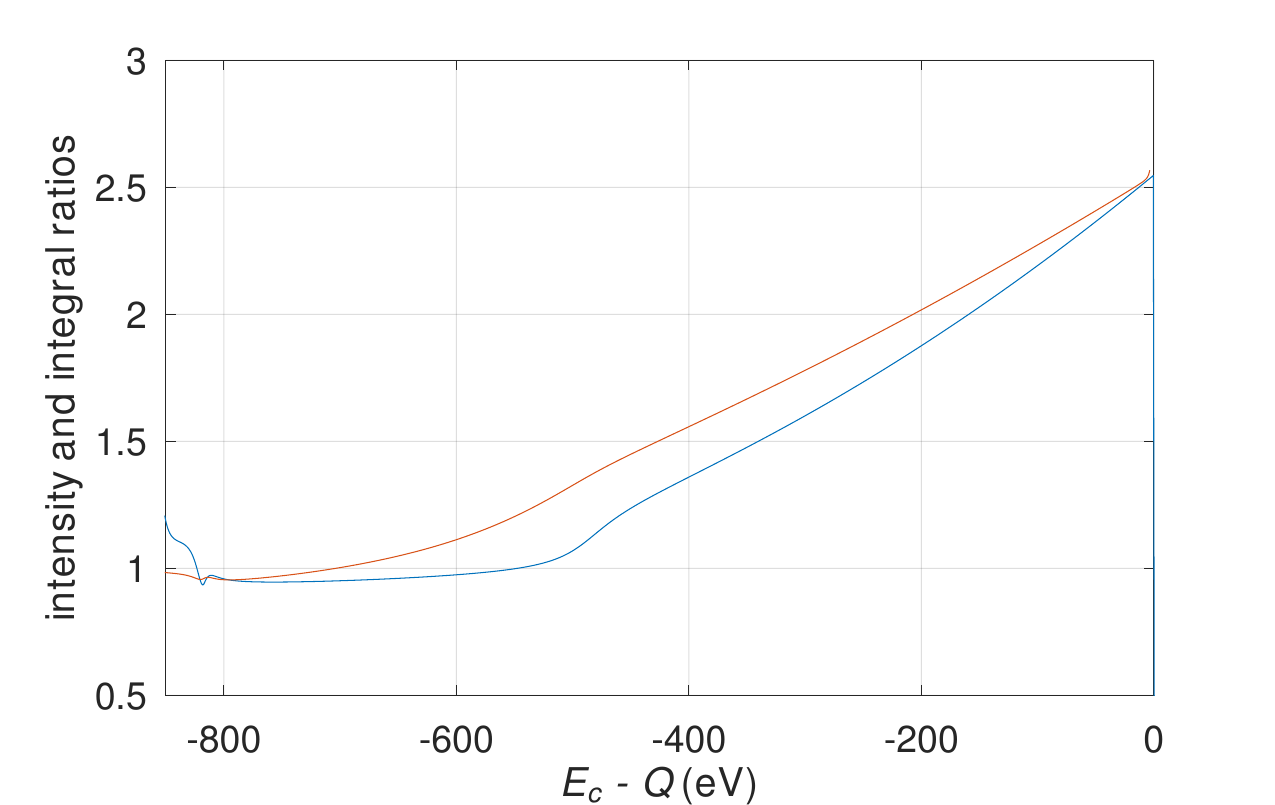}
\caption{Ratio of the normalized spectra (blue) and the ratio of their integrals (red) calculated with the single-hole and phenomenological models below the end-point. The abscissa shows $E_c - Q$, i.e., the distance from the endpoint, and the integrals entering the ratio are numerically computed from $E_c$ to $Q$.}
\label{fig:endpoint-cfr}
\end{figure*}

Of particular relevance to holmium-based neutrino mass experiments is the M1N2 shake-off,
\footnote{\label{foot:M1N2tail} 
For $E_b$ in this shake-off spectrum the fit yielded an extremely large, effectively divergent value (\emph{unbound} in Table\,\ref{tab:fit_params_shakeoff}). This reflects 
the total insensitivity of the fit to this parameter. This is expected, as the spectral shape is almost entirely determined by the phase space factor. Consequently, the integral of this shake-off spectrum does not carry physical meaning.}
which contributes to the enhanced spectral intensity near the end-point, together with the pronounced asymmetry of the M1 resonance. These effects increase the expected neutrino mass sensitivity compared to the simplified single-hole spectrum assumed in \cite{derujula_calorimetric_1982}.
See Fig.\,\ref{fig:endpoint-cfr} for the ratio of the relative count rates between the single-hole spectrum and our phenomenological model above the M1 resonance.
It is also noteworthy that the successful {\it ab initio} approach does not predict any fine structure in the end-point region that could affect the neutrino mass measurement. 

To assess the ability of the phenomenological model to correctly describe the spectral shape above the M1 resonances we repeated the analysis presented in \cite{alpert_most_2025} using the functional form given by Eq.\,\ref{eq:sof} for the M1N2 shake-off excitation and replacing the low degree polynomial $\mathcal{S}_\mathrm{pol}$ with all the other components of the phenomenological model.
The parameters listed in Table\,\ref{tab:e0data} are used to define informative priors for the Bayesian fit. Applying the Bayesian analysis over the same energy interval yields an upper limit on the electron neutrino mass of $m_{\nu}<31$\,eV$/c^2$ at 90\% credibility, and an end-point energy $Q = (2851\pm6)$\,eV, in excellent agreement with the results reported in \cite{alpert_most_2025}. 
A detailed discussion of the neutrino mass extraction procedure and the associated systematic uncertainties falls outside the scope of the present work, which focuses on the phenomenological decomposition of the entire calorimetric spectrum. A comprehensive analysis of systematic effects in the end-point region is presented in \cite{alpert_most_2025}.

The last column in Table\,\ref{tab:e0data} reports the intensity of each component, expressed as the ratio of its numerical integral to that of the M1 peak.
The integrals are computed numerically for the functions in Eqs.\,\ref{eqn:bw} and \ref{eq:sof} over the interval from 0\,eV to three times the $Q$ value.
For comparison with theoretical predictions, Table\,\ref{tab:rel_intensities} reports the capture probabilities for the shells directly accessible via $^{163}$Ho electron capture ($n_iC_i\beta_i^2B_i$ in Eq.\,\ref{eq:E_c-distr}). These values are normalized to the M1 probability and are obtained by summing the relative intensities of all excitations involving the same core hole (H1).
Excitations with core holes in shells not directly accessible via EC are not included in the sum, since it is not possible to establish from which original core hole they have been scattered.
\begin{table}[htbp]
  \centering
  \caption{Electron capture probabilities ($n_iC_i\beta_i^2B_i$ in Eq.\,\ref{eq:E_c-distr}) for the directly accessible shells normalized to M1. See text for the details about the calculation of the capture ratios. The probability quoted for the O2 line strongly depends on the chosen energy position (see footnote on pag.\,\pageref{foot:O2}).}
  \label{tab:rel_intensities}
  \begin{tabular}{|c|c|}
    \hline
    Shell & Probability \\
     & (relative to M1) \\
    \hline\hline
    O2 & 0.0022(1) \\
    O1 & 0.0386(9) \\
    N2 & 0.0109(1) \\
    N1 & 0.228(2) \\
    M2 & 0.0615(1) \\
    \hline
  \end{tabular}
\end{table}

\section{Conclusions}
With this work, we have provided a phenomenological decomposition of the calorimetric decay spectrum of $^{163}$Ho measured by the HOLMES experiment.
The unfolded spectrum is reproduced by a combination of Breit-Wigner resonances and shake-off spectra, for which we provide all parameters.
Additionally, we have given a tentative interpretation of all components and compared our phenomenological model with the results of {\it ab initio} calculations reported in \cite{brass_initio_2018,brass_initio_2020}.
We find that, despite the limited accuracy of the {\it ab initio} results---due to numerical approximations and limitations---and the low statistics of our experimental spectrum, the agreement is fairly good.
Our model confirms many predictions not present in the works of other authors.

The phenomenological model presented in this work has a strong impact on future neutrino mass measurements planning to use $^{163}$Ho EC.
The model provides a precise description of the endpoint region, which can be used to analyse experimental data in terms of neutrino mass, including backgrounds due to both the pile-up spectrum and the tails of all components with energies below the region of interest.
In addition, the modelling, extending from as low as about 30\,eV up to $Q$, allows accurate Monte Carlo generation of toy spectra needed to assess the sensitivity of future experiments.
Last but not least, the detailed decomposition of the calorimetric spectrum in terms of atomic de-excitations provides a tool for assessing the sequence of single energy depositions contributing to each calorimetric energy detection.
This is a critical ingredient for future studies of possible systematic uncertainties in the neutrino mass measurement due to solid-state and detector effects.
\section*{Acknowledgements}
The authors gratefully acknowledge Alvaro De R\'ujula, Martin Brass, and Jonathan Dean for insightful discussions, and thank Martin Brass for providing numerical data from his publications. The HOLMES experiment is supported by the Istituto Nazionale di Fisica Nucleare (INFN) and by the European Research Council under the European Union’s Seventh Framework Programme (FP7/2007–2013), ERC Grant Agreement no. 340321.
\clearpage
\appendix
\addcontentsline{toc}{section}{Appendices}

\section*{Appendices}

\section{Parameters for the fitted components}
\label{app:tables}
\begin{table*}[htbp]
\centering
\caption{Fitted parameters for the Breit-Wigner peaks. Values are reported as mean (standard error). The standard errors include both statistical and systematic uncertainties; the systematic component reflects only the propagation of uncertainties from the unfolded spectrum. The parameters for the first peak are not obtained from a fit and therefore have no associated uncertainty (see Sec.\,\ref{ssec:fit} for details).}
\label{tab:fit_params_asym_bw}
\resizebox{\textwidth}{!}{
\begin{tabular}{|c|c|c|c|c|c|c|c|}
\hline
ID & $E_0$ & $A^{\mathrm{BW}}$ & $\Gamma$ & $\delta_{AS}$ & $E_{th}$ & $p$\\
& (eV) & (count)$\times 10^9$ & (eV) & & (eV) & \\
\hline
\hline
0    & 29(0) & 6.0(0) & 2.5(0) & -- & -- & -- \\
\hline
1 & 41.7(8) & 0.22(2) & 7.9(5)  & -- & -- & -- \\
2  & 50.05(4) & 30.9(3) & 6.90(7) & 0.68(1) & 44.36(4) & 1.24(4) \\
3 & 60.54(6) & 3.02(5) & 7.3(1) & -- & -- & -- \\
4 & 312.9(2) & 0.17(1) & 8.8(4) & -- & -- & -- \\
5  & 329.22(8) & 7.08(9) & 15.4(2) & 0.68(2) & 309.7(3) & 0.27(3) \\
6 & 339.93(9) & 1.45(3) & 14.5(2) & -- & -- & -- \\
7  & 409.65(8) & 74.4(4) & 7.63(3) & 0.618(4) & 0.000(1) & $7(5)\times 10^{-6}$\\
8  & 412.9(1) & 297(3) & 21.5(2) & 0.145(2) & 394.9(3) & 3.22(6) \\
9  & 416.4(2) & 24.2(6) & 4.20(8) & 0.62(1) & 0.000(1) & $6(5)\times 10^{-4}$\\
10 & 456.15(3) & 5.54(2) & 18.2(1) & -- & -- & -- \\
11 & 498.1(4) & 0.55(5) & 22(1) & -- & -- & -- \\
12 & 509.9(3) & 0.29(3) & 16.4(8) & -- & -- & -- \\
13 & 1287.3(7) & 0.0026(4) & 5(2) & -- & -- & -- \\
14 & 1330.6(5) & 0.0068(7) & 10(1) & -- & -- & -- \\
15  & 1621.0(7) & 0.0052(7) & 7(1) & -- & -- & -- \\
16  & 1718.8(5) & 0.017(1) & 10(1) & -- & -- & -- \\
17  & 1836.25(3) & 39.63(3) & 8.513(6) & 0.932(1) & 0.000(1) & $1.0(1)\times 10^{-7}$\\
18  & 2000.66(7) & 3.68(3) & 12.48(7) & -- & -- & -- \\
19  & 2041.25(2) & 794.5(2) & 14.383(6) & 0.8763(7) & 1499.1(1.2) & 1.424(5) \\
\hline
\end{tabular}
}
\end{table*}
\begin{table*}[htbp]
\centering
\caption{Fitted parameters for shake-off spectra.  Values are reported as mean (standard error). The standard errors include both statistical and systematic uncertainties; the systematic component reflects only the propagation of uncertainties from the unfolded spectrum. See footnote on page \pageref{foot:M1N2tail} for a comment on the $E_b$ found for the last shake-off in the table.}
\label{tab:fit_params_shakeoff}
\begin{tabular}{|c|c|c|c|c|c|}
\hline
ID & $E_0$  & $A^{\mathrm{SOF}}$ & $\Gamma$ & $E_b$  \\
& (eV) & (count)$\times 10^9$ & (eV) &  (eV) \\
\hline \hline
1  & 58.45(6) & 0.416(4) & 0.004(3) & 364(5) \\
2  & 479.6(1) & 3.57(6) & 10.9(3) & 92(1) \\
3  & 573.4(7) & 0.25(1) & 21(3) & 282(19) \\
4  & 757(3) & 0.208(8) & 164(12) & 806(22) \\
5  & 1406(6) & 0.046(2) & 142(13) & 1233(144) \\
6  & 2371(2) & 0.1048(6) & 57(5) & \emph{unbound} \\
\hline
\end{tabular}
\end{table*}

\clearpage

\section{Fit parameter correlations}
\label{app:corr}
\begin{figure*}[h!]
  \includegraphics[width=0.99\textwidth]{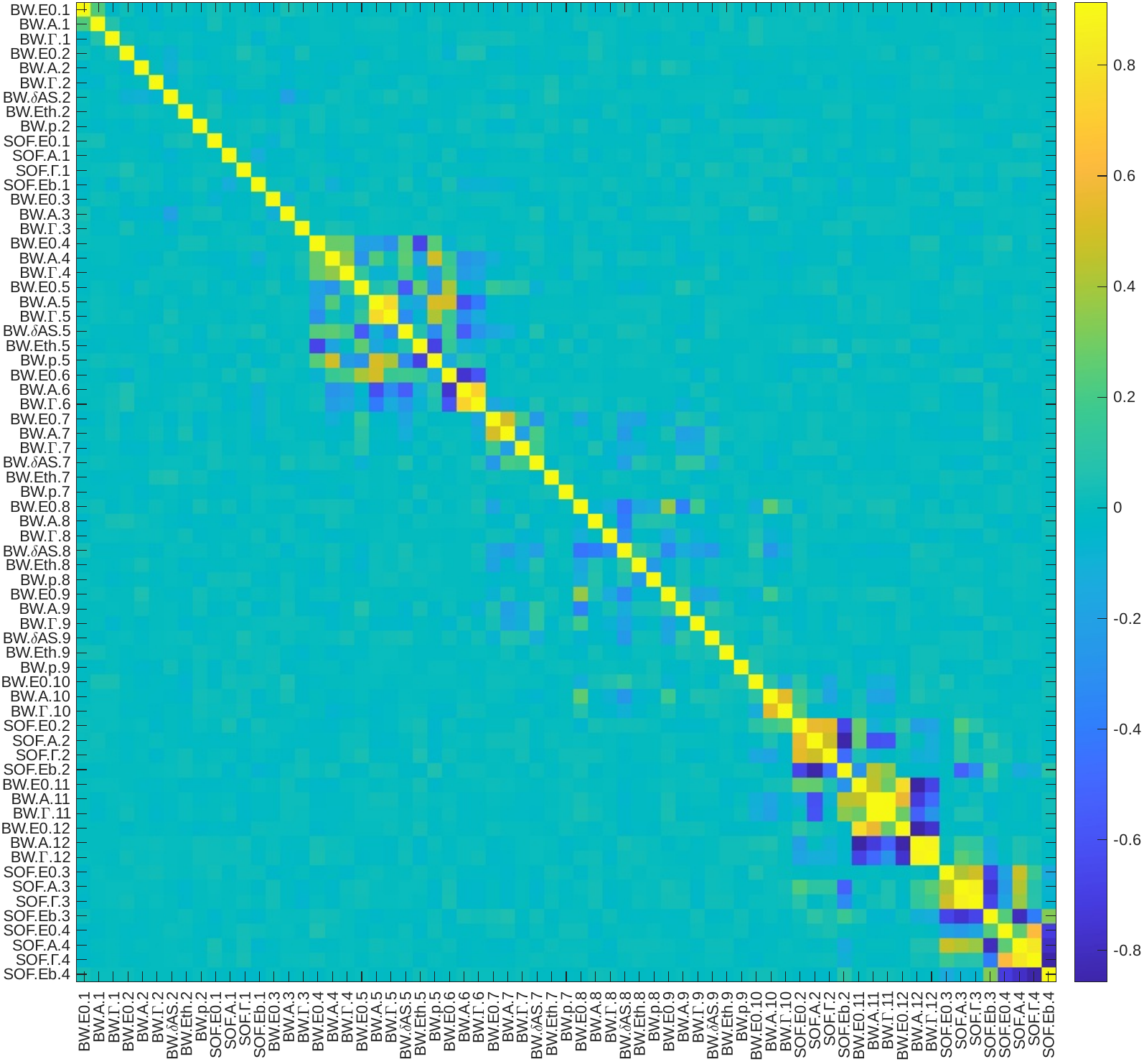}
    \caption{Correlations between the fit parameters of the O and N group components below 1\,keV (see Fig.\,\ref{fig:NO_bare}). Both Breit-Wigner (BW) and shake-off (SOF) components are ordered by increasing energy position $E_0$. The parameter numbering correspond to the IDs in Tables\,\ref{tab:fit_params_asym_bw} and \ref{tab:fit_params_shakeoff}. 
    }
\label{fig:ONcorr}
\end{figure*}

\begin{figure*}[h!]
  \includegraphics[width=0.99\textwidth]{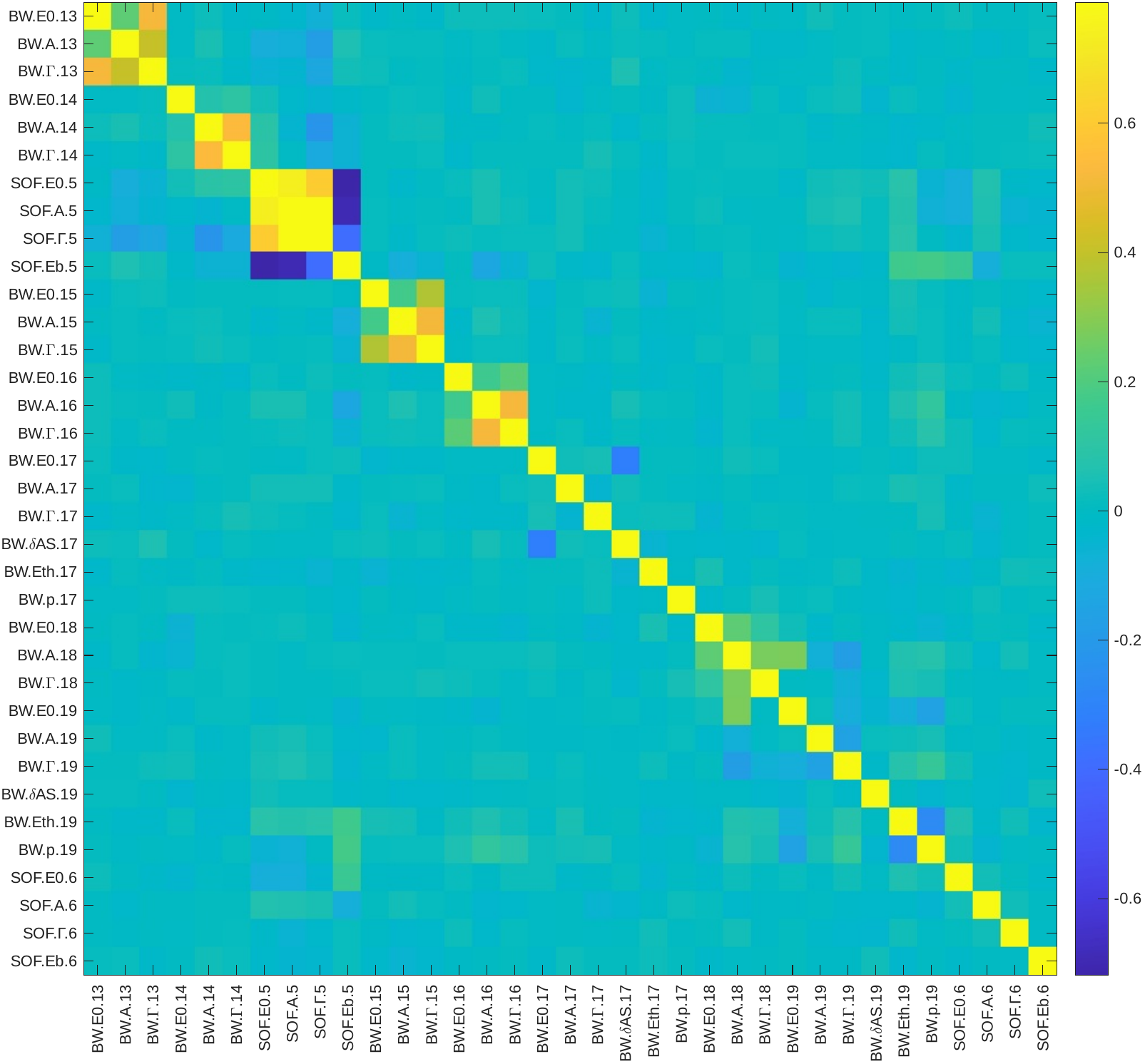}
  \caption{Correlations between the fit parameters of the M group components above 1\,keV (see Fig.\,\ref{fig:full_bare}). Both Breit-Wigner (BW) and shake-off (SOF) components are ordered by increasing energy position $E_0$. The parameter numbering correspond to the IDs in Tables\,\ref{tab:fit_params_asym_bw} and \ref{tab:fit_params_shakeoff}. }
\label{fig:Mcorr}
\end{figure*}
\clearpage

\providecommand{\href}[2]{#2}\begingroup\raggedright\endgroup

\end{document}